\documentclass[
  amssymb, amsmath,
  floatfix,
  reprint,%
  twocolumn,%
  superscriptaddress,%
]{revtex4-1}

\usepackage{graphicx}
\usepackage{caption}

\usepackage{hyperref}
\hypersetup{
     colorlinks=true,
     linkcolor=blue
}

\usepackage[T1]{fontenc} 
\usepackage[utf8]{inputenc}
\usepackage{lmodern}

\usepackage{booktabs}
\usepackage{etoolbox}
\usepackage[separate-uncertainty=true]{siunitx}
\DeclareSIUnit\wn{\per\cm}
\usepackage{upgreek}
\usepackage{chemmacros}
\chemsetup{
 formula = mhchem,
 greek = chemgreek,
 modules = thermodynamics,
 modules = redox,
 modules = reactions
}
\usepackage[symbol]{footmisc}
\renewrobustcmd{\bfseries}{\fontseries{b}\selectfont}
\renewrobustcmd{\boldmath}{}

\usepackage{booktabs, dcolumn} 
\newcolumntype{d}[1]{D{.}{.}{#1}}

\begin{document}
\title{Conformational Landscape of Oxygen-Containing Naphthalene Derivatives}

\author{Manuel Goubet}
\affiliation{Universit\'e Lille, CNRS, UMR 8523-PhLAM-Physique des Lasers, Atomes et Mol\'ecules, F-59000 Lille, France}
\email{manuel.goubet@univ-lille.fr}

\author{Marie-Aline Martin-Drumel}
\affiliation{Universit\'e Paris-Saclay, CNRS, Institut des Sciences Mol\'eculaires d'Orsay, 91405 Orsay, France}

\author{Florent R\'eal}
\affiliation{Universit\'e Lille, CNRS, UMR 8523-PhLAM-Physique des Lasers, Atomes et Mol\'ecules, F-59000 Lille, France}

\author{Val\'erie Vallet}
\affiliation{Universit\'e Lille, CNRS, UMR 8523-PhLAM-Physique des Lasers, Atomes et Mol\'ecules, F-59000 Lille, France}

\author{Olivier Pirali}
\email{olivier.pirali@universite-paris-saclay.fr}
\affiliation{Universit\'e Paris-Saclay, CNRS, Institut des Sciences Mol\'eculaires d'Orsay, 91405 Orsay, France}

\begin{abstract}
Polycyclic aromatic compounds (PACs) constitute an important class of molecules found in various environments and are considered important pollutants of the Earth's atmosphere. In particular, functionalization of PACs modify the ring aromaticity, which greatly influences the chemical reactivity of these species. In this work we studied several oxygen-containing PACs, relevant to atmospheric chemistry. We investigated the conformational landscape of four naphthalene-derivative molecules --- namely ,1- and 2-hydroxynaphthalene and 1- and 2-naphthaldehyde ---  by means of rotational and vibrational spectroscopy supported by quantum chemical calculations. For 1-hydroxynaphthalene and 1-naphthaldehyde, intramolecular hydrogen bonding and steric effects drive the conformational preferences while for  2-hydroxynaphthalene and 2-naphthaldehyde,  the charge distributions allow us to understand the conformational landscape.
This work not only demonstrates how the localization of the substitution group in the ring influences the conformational relative energies and but also constitutes a step toward a better understanding of the different chemical reactivity of such functionalized PACs.
\end{abstract}
\maketitle

\section{Introduction}
\label{sec:intro}

Polycyclic aromatic compounds (PACs), which represent about two-thirds of the so far known chemical compounds~\cite{Balaban04}, are of high relevance in diverse fields ranging from astrophysics and atmospheric chemistry to combustion science. Within this very rich class of molecules, purely carbonaceous species (also called polycyclic aromatic hydrocarbons, PAHs) may be those that have received the most attention in the laboratory so far, in part driven by astrophysical motivations (see, \textit{e.g.}, the review by~\citet{Tielens08}). Indeed, since the early hypothesis formulated by~\citet{Leger84} and~\citet{Allamandola85}, PAH molecules are suspected to be abundantly present in the interstellar medium. Their neutral, ionized, and radical forms have been suggested as carriers of many broad absorption and emission features observed in various objects of the universe (see, \textit{e.g.}, ref~\citenum{Salama2008}). Besides PAHs, functionalized variants are relevant to many scientific fields as well. Recently, several pure rotational transitions of a functionalized aromatic compound, benzonitrile (\ce{C6H5-CN}), have been unambiguously detected in the spectrum of the TMC-1 molecular cloud~\cite{McGuireBenzonitrile2018}, confirming the presence of PACs in the interstellar medium.\par
PAC molecules also constitute an important class of pollutants of the Earth's atmosphere; they originate mainly from incomplete combustion processes (\textit{e.g.}, residential heating, vehicle exhausts) and can be found both in gas and in particle phases~\cite{Ravindra2008}. Functionalized species can either be direct products of such combustion processes or secondary products resulting from oxidation of primary PAH species. Recent studies that aimed at measuring the seasonal variations in the abundances of more than 80 PACs in an urban environment highlighted both the importance of such species for Earth's atmosphere science and the concern raised by their influence on atmospheric chemistry and, to some extent, human health~\cite{TomazPACsGrenoble2016}. In the atmosphere, oxygen-containing PACs are almost as abundant as PAHs while their vapor pressures are about 10 times lower than their corresponding parent PAH species and therefore constitute an important class of aerosol precursors~\cite{KeyteReview2013,Yuan2013}. Despite their relevance to Earth's atmosphere, very little is known concerning the spectroscopy and structural information on these compounds, preventing both their efficient tracking and a better understanding of their formation/destruction chemical networks.\par
Aside from the importance of PACs in atmospheric science and astrophysics, fundamental questions regarding the influence of a functional group on aromaticity remain open. Indeed, while PAHs are relatively chemically stable compounds, when functionalized their reactivity either increases or decreases drastically. 
Substitution of a hydrogen by a functional group, either electron-withdrawing or electron donor, obviously breaks the symmetric electronic distribution over the ring, modifying its affinity for electrophilic attack. For instance, PACs containing hydroxyl or amino groups (activating groups) are more reactive than the corresponding, nonfunctionalized, PAHs. Conversely, PACs containing aldehyde or nitro groups (deactivating groups) are less reactive than their parent PAHs. In addition to their influence on chemical reactivity, functionalized PACs tend to be more easily adsorbed than their PAHs counterparts~\cite{Huffer2017}, revealing an increased affinity for long-range polar interactions. At the molecular level, the addition of functional groups decreases the degree of symmetry of most PACs. If the functional group is itself asymmetric, at least two conformers (distinct by a 180\textdegree\ rotation around the bond linking the functional group to the PAHs fragment) will coexist but will not be isoenergetic for reasons that deserve elucidation. Such conformational energy differences are particularly small and can only be captured by highly accurate quantum chemical calculations.\par
In this work, pure rotation spectroscopy in the centimeter and millimeter wave regions together with vibrational spectroscopy in the infrared (IR) range have been used to characterize four oxygen containing (two hydroxyls and two aldehydes) functionalized naphthalene molecules, namely, 1- and 2-hydroxynaphthalene (\ce{C10H7-OH}), and 1- and 2-naphthaldehyde (\ce{C10H7-C(O)H}). Electronic structure calculations have been undertaken not only to support the spectral analyses but also to elucidate the conformational preferences. The present study provides valuable data enabling the confident search for these species in the Earth's atmosphere and interstellar medium. Finally, in light of the experimental and theoretical results, the stability of the different isomers and conformers is discussed in terms of asymmetric charges distribution, providing some insights into the influence of functional groups on aromaticity at the molecular level.

\section{Laboratory Methods}

\subsection{Quantum Chemical Calculations}
\begin{figure*}
	\includegraphics[width=\textwidth]{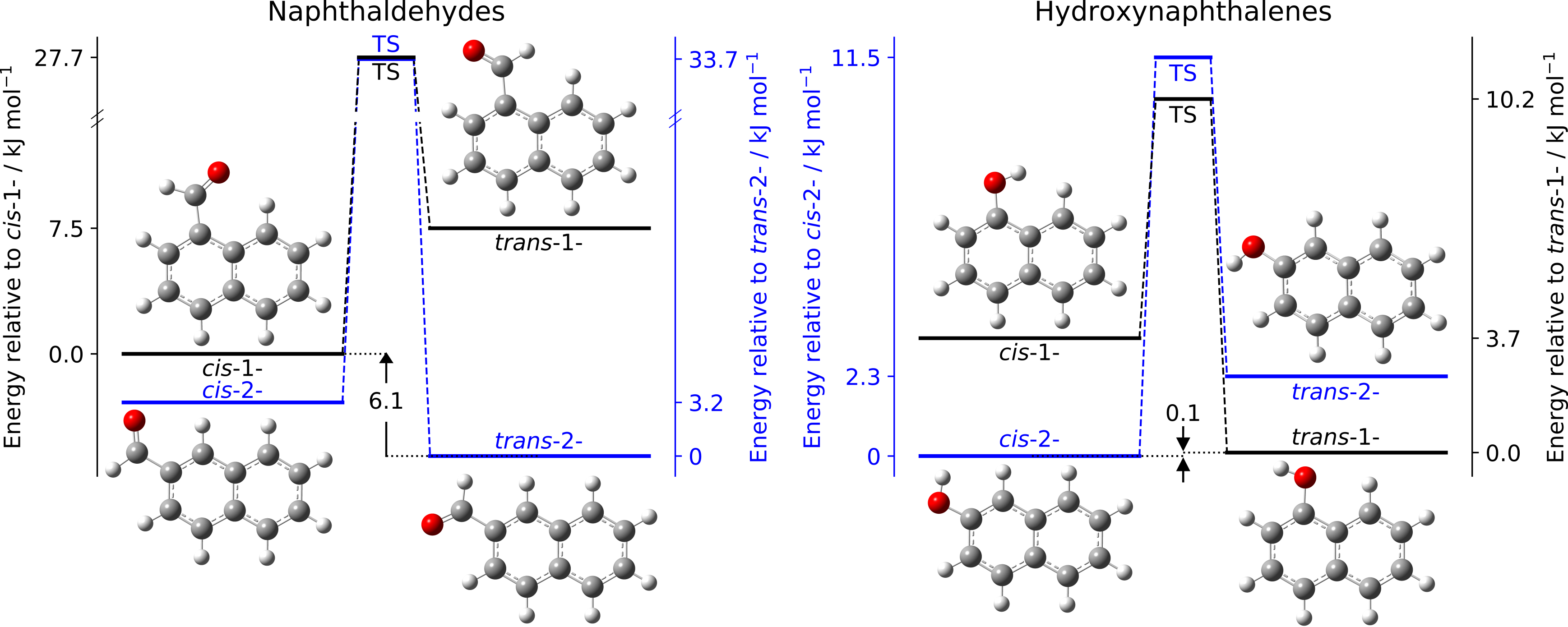}
\caption{Schematic view of the conformational landscapes of the naphthaldehyde (\textit{left trace}) and hydroxynaphthalene (\textit{right trace}) compounds. The isomers of lower energy, $i.e.$, 2-naphthaldehyde and 2-hydroxynaphthalene, are plotted in blue and the energy difference between the 1- and 2- compounds is reported (in \si{\kJ\per\mol}). Relative zero point energies have been calculated by using the hybrid method described in the text.}
	\label{fig:ConformLand}
\end{figure*}

\subsubsection{Structures and Zero Point Energies}
All the calculations were performed using Gaussian 16 rev B.01 software~\cite{g16} on the PhLAM laboratory computing cluster. The equilibrium geometries of each of the two stable conformers (\textit{cis} and \textit{trans}, Figure~\ref{fig:ConformLand}) of the studied compound were fully optimized both using second order M{\o}ller-Plesset perturbation theory (MP2) and density functional theory (B97-1) with Dunning's correlation consistent basis sets (aug-)cc-pVXZ (X=D,T,Q)~\cite{dunning_gaussian_1989} using the tight convergence option. Following these optimizations, harmonic frequencies calculations were performed using a tight SCF convergence criterion, the ultrafine integral grid option, and the ``vibrot'' option to obtain harmonic vibrational frequencies --- (thus zero-point energy (ZPE) correction to the energies) --- and the vibration-rotation coupling (quartic centrifugal distortion) constants.

\subsubsection{Vibrational Frequencies and Rotational Constants}
Anharmonic vibrational frequencies and rotational constants in the ground and vibrationally excited states have been estimated from a ``hybrid'' method~\cite{Barone10,Grimme13}, recently used with success on other oxygen-containing species~\cite{Goubet15,Bteich18}.
Briefly, since vibrational second-order perturbation theory (VPT2) calculations at the MP2 level as implemented in Gaussian 16 are not affordable for relatively large molecules as those of interest here, ``hybrid'' rotational constants of a given vibrational state $v$ are estimated by adding the DFT anharmonicity corrections (B97-1/cc-pVTZ) to the MP2 equilibrium constants (MP2/cc-pVTZ). Once the experimental constants in the ground vibrational state (GS) are known, the predicted constants in vibrationally excited states (ES) can further be corrected to take into account the vibrational GS errors, a method that has proven useful in the past to support the analyses of the vibrational ES spectra of many molecules, as for instance PAHs~\cite{Gruet14}. This correction was particularly useful to assign the pure rotation lines involving ES of $trans$-1-hydroxynaphthalene as shown further in the text.

\subsubsection{Isomeric and Conformational Landscapes}

For 1-hydroxynaphthalene, a MP2/cc-pVTZ potential energy curve (PEC) along the \ce{C-O} torsional angle has been calculated by means of a relaxed scan procedure along the \ce{C-C-O-H} dihedral angle with 5\textdegree\ steps. The resulting PEC permitted us to ensure that the torsion is a very classical nonequivalent 2-fold potential with the functional group in the plane of naphthalene at equilibria and perpendicular to the plane at transition states (TS),  TS mirror images being strictly equivalent in energy (see Figure~S1 in the Supporting Information). Therefore, only TS searches (using the QST method as implemented in Gaussian 16) have been performed for the three other compounds. The relative ZPE corrected energies of each conformer and the height of the barriers between them (\textit{i.e.}, torsion of the alcohol or aldehyde group) have been estimated at the ``hybrid'' level described above and are presented in Figure \ref{fig:ConformLand} and Tables~S1 and S2 in the Supporting Information.

Since the computed relative energies between isomers/conformers are rather small (few \si{\kJ\per\mol} in most cases), it is relevant to estimate the uncertainty relative to the choice of both the correlated electronic structure method and the size of the atomic basis sets. In quantum chemistry, the most accurate estimation of the dynamic correlation is achieved by using complete basis set (CBS) extrapolation and the gold standard coupled-cluster method with iterative inclusion of single and double excitations and perturbative inclusion of triple excitations (CCSD(T)). In the case of  functionalized naphthalene compounds, the precision of the hybrid method can be estimated from single-point calculations of the relative energies. Such calculations have been performed on naphthaldehyde compounds with the MOLPRO 2019 package~\cite{MOLPRO_brief} using various wave function methods [MP2, CCSD(T)] extrapolated to the CBS limit, as well as localized explicit-correlated method PNO-LCCSD(T)-F12 (PNO being pair natural orbitals)~\cite{Ma2018,Werner2017}. The results of this comparison (see Table~S3 in the Supporting Information) show that no notable differences in relative energies are obtained compared to the hybrid level (changes by few \% only). Thus, for all other compounds, the discussion is based on the results of the hybrid method only. 

\subsubsection{Charges Distributions and Molecular Orbitals}
The natural bond orbital (NBO) analysis program (version 6.0)~\cite{NBO6} has been used to compute the atomic charge distribution and pinpoint the presence of intramolecular hydrogen bonds. Indeed, NBO transforms the delocalized, many-electron wave function into optimized electron pair bonding subunits, i.e., in a set of Lewis-type (such as bonds and lone pairs) and non-Lewis-type (such as Rydberg and antibond) NBOs. Since a hydrogen bond can be viewed as a charge transfer from a donor (oxygen lone pair) to an acceptor (antibonding \ce{C-H}), its strength can be quantified by the stabilization energy E(2) computed at the second-order perturbation-theory by the NBO analysis, when larger than the default threshold \SI{2}{\kJ\per\mol}. Naphthaldehyde isomers were used to test the variations caused by various method/basis set associations (MP2 and B97-1 methods, cc-pVXZ and aug-cc-pVXZ basis sets, with X = D,T,Q). Absolute values of atomic charges are notably changing according to the method/basis set used. However, all tendencies discussed hereafter from results at the MP2/aug-cc-pVQZ level (considered as the highest one) stay the same, from a relative point of view, whatever the method used. NBO calculations have been subsequently performed on hydroxynaphthalene, naphthalene, benzaldehyde, and hydroxybenzene.

\subsection{Chemical Samples}

All four samples, 1- and 2-hydroxynaphthalene and 1- and 2-naphthaldehyde, have been purchased from Aldrich and used without further purification. 

\subsection{Centimeter Wave Spectroscopy}

\begin{figure}[htbp]
	\includegraphics[width=\columnwidth]{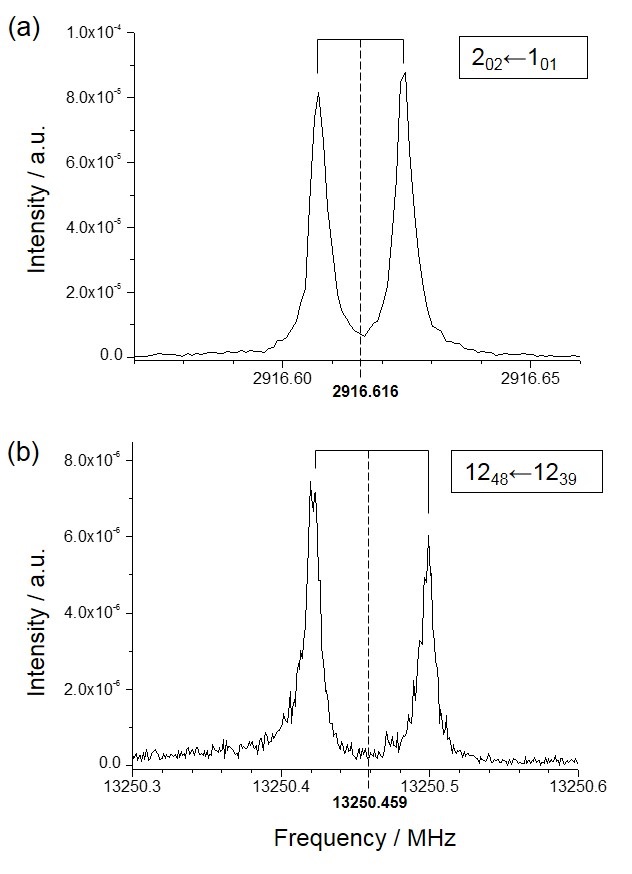}
	\caption{Examples of FTMW lines (in the shape of Doppler doublets resulting from the coaxial arrangement of the cavity and the supersonic expansion) of 2-hydroxynaphthalene: (a) strong $^\mathrm{q}$R-type and (b) weak $^\mathrm{p}$Q-type transitions of the \textit{cis} conformer. These spectra correspond to 100 and 1000 acquisitions, respectively. Assignments are in the format $J'_{K'_{a}K'_{c}} \leftarrow J"_{K"_{a}K"_{c}}$. Center frequency for each doublet is taken as the average of the two Doppler components.}
	\label{fig:FTMW}
\end{figure}

The pure rotation spectra of 1- and 2-hydroxynaphthalene and 1- and 2-naphthaldehyde have been recorded in the 2--\SI{20}{\GHz} frequency range using a supersonic jet Fourier transform microwave (FTMW) spectrometer located at the PhLAM Laboratory (see refs~\citenum{Kassi2000,Tudorie2011} for details).  A heated nozzle~\cite{Kassi2004} allowed to mix the sample vapor with the carrier gas (neon) at a backing pressure of about \SI{0.3}{\MPa}. Temperatures ranging from 393 to \SI{413}{\kelvin} were found to optimize the signal-to-noise ratio (SNR) depending on the amplitude of the recorded lines. The spectral resolution was set to 1.8 kHz, subsequently considered as the line accuracy. Examples of observed signals are displayed in Figure~\ref{fig:FTMW}.

\subsection{Millimeter Wave Spectroscopy}

Room temperature millimeter-wave absorption spectra of each of the four species were recorded at ISMO laboratory. The spectrometer, described in~\citet{PiraliAzaAda2017}, is composed of a radio frequency synthesizer driving a commercial frequency multiplier source from Virginia Diodes Inc. In the present study, all spectra were recorded in the 140--\SI{220}{\GHz} range using similar experimental conditions: a \SI{49}{\kHz} frequency modulation of the signal with a modulation depth of \SI{300}{\kHz}, a \SI{25}{\ms} time constant, 30 kHz frequency steps, and a second harmonic lock-in detection were used. The collimated radiation was injected in a \SI{2.25}{\m} absorption cell equipped with Teflon windows. The output radiation was focused onto a Schottky diode detector from Virginia Diodes Inc. A constant flow of sample at a pressure lower than \SI{0.1}{\Pa} was used to record the spectra. Under these experimental conditions, the line center frequency accuracy ranges from 50 to \SI{75}{\kHz}, based on the SNR.
\begin{figure*}
	\includegraphics[width=\textwidth]{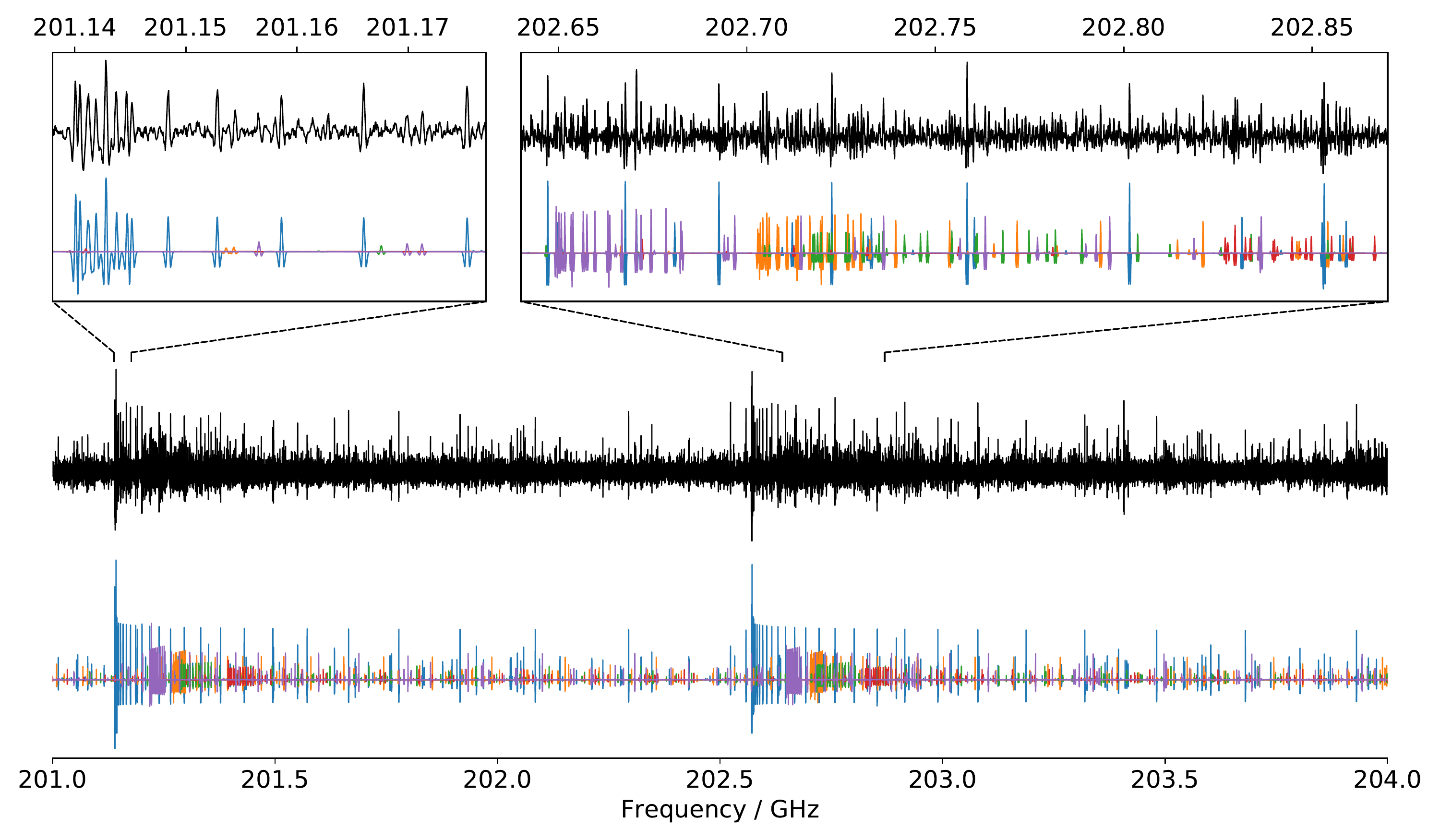}
\caption{Portions of the room temperature experimental millimeter spectrum of 1-hydroxynaphthalene (top traces, in black) and comparison with a simulation (bottom traces) of the pure rotational transitions within the ground state (in blue) and within several low-lying vibrational excited states (various colors; see the Supporting Information). The broader portion of the spectrum displays two consecutive $^\mathrm{q}\mathrm{R}$ series with $J''+K''_a=139/140$ and 140/141 (unresolved asymmetric splitting). The PGOPHER software was used to simulate the spectrum at \SI{300}{\kelvin} using the rotational constants in each state derived in this work. Relative intensities of the ES rotational lines are modeled using the calculated (hybrid method) energies of each mode and a Boltzmann distribution at \SI{300}{\kelvin}. The second derivative of the obtained traces was subsequently calculated to allow better comparison with the experimental spectrum.}
	\label{fig:mmw1Naphtol}
\end{figure*}

Figure~\ref{fig:mmw1Naphtol} shows a small portion of the 1-hydroxynaphthalene absorption spectrum. Several clusters of lines corresponding to $^\mathrm{q}$R-type transitions within GS and several ES are clearly identified. The simulated spectrum is obtained using the final set of rotational constants as given in Tables~\ref{tab:RotationNaphtols} and S6 in the Supporting Information.

\subsection{Fourier-Transform IR Spectroscopy} \label{sec:FTIR}
In addition to the pure rotation measurements, the vibrational spectra of the two naphthaldehyde species were recorded in the 50--\SI{3500}{\wn}  spectral range. The absorption spectra were recorded using the Bruker IFS125HR Fourier-transform (FT) interferometer located on the AILES beamline of the SOLEIL synchrotron. Room temperature vapor pressure of each sample (typically of the order of a few Pa) was injected in a White-type multipass cell with optics set to obtain 150 m path length~\cite{Martin-Drumel13}. We attempted to record the absorption spectra of the two hydroxynaphthalene molecules in the same conditions but their lower vapor pressure prevented from any vibrational band detection. For the naphthaldehydes spectra, the rotational structures of the vibrational bands could not be resolved at the ultimate resolution of the FT interferometer (\SI{0.001}{\wn}), we chose a spectral resolution of \SI{0.5}{\wn} to record the vibrational signatures. Since such relatively low spectral resolution does not require the use of synchrotron radiation as a continuum source for the interferometer, a globar source was used instead. A \SI{6}{\micro\meter} mylar beamsplitter and a liquid He cooled bolometer were used to record the spectra in the 50--\SI{600}{\wn} range. For the 600--\SI{3500}{\wn} range, a KBr beamsplitter was used together with a liquid nitrogen cooled HgCdTe detector.

\section{Results}
\label{sec:results}

The quantum chemical relative energies of the conformers allow to evaluate their population at \SI{300}{\kelvin}. For 1-naphthaldehyde, 95\,\% of the population is in the \textit{cis} form while for 2-naphthaldehyde 79\,\% of the population is in the \textit{trans} form. Concerning 1-hydroxynaphthalene, 82\,\% of the population is in the \textit{trans} form while for 2-hydroxynaphthalene, 71\,\% is in the \textit{cis} form (see Tables S1 and S2 in the Supporting Information).

\subsection{Vibrational Spectra of 1- and 2-Naphthaldehyde}

All four naphthaldehyde compounds ($cis$ and $trans$ of both 1- and 2-species) belong to the $C_\mathrm{s}$ symmetry point group and possess 37\,A$'$ and 17\,A$''$ vibrational modes. Vibrational bands of A$'$ symmetry are of $a/b$-type while those of A$''$ symmetry are of $c$-type.

\begin{figure}[htbp]
	\includegraphics[width=0.9\columnwidth]{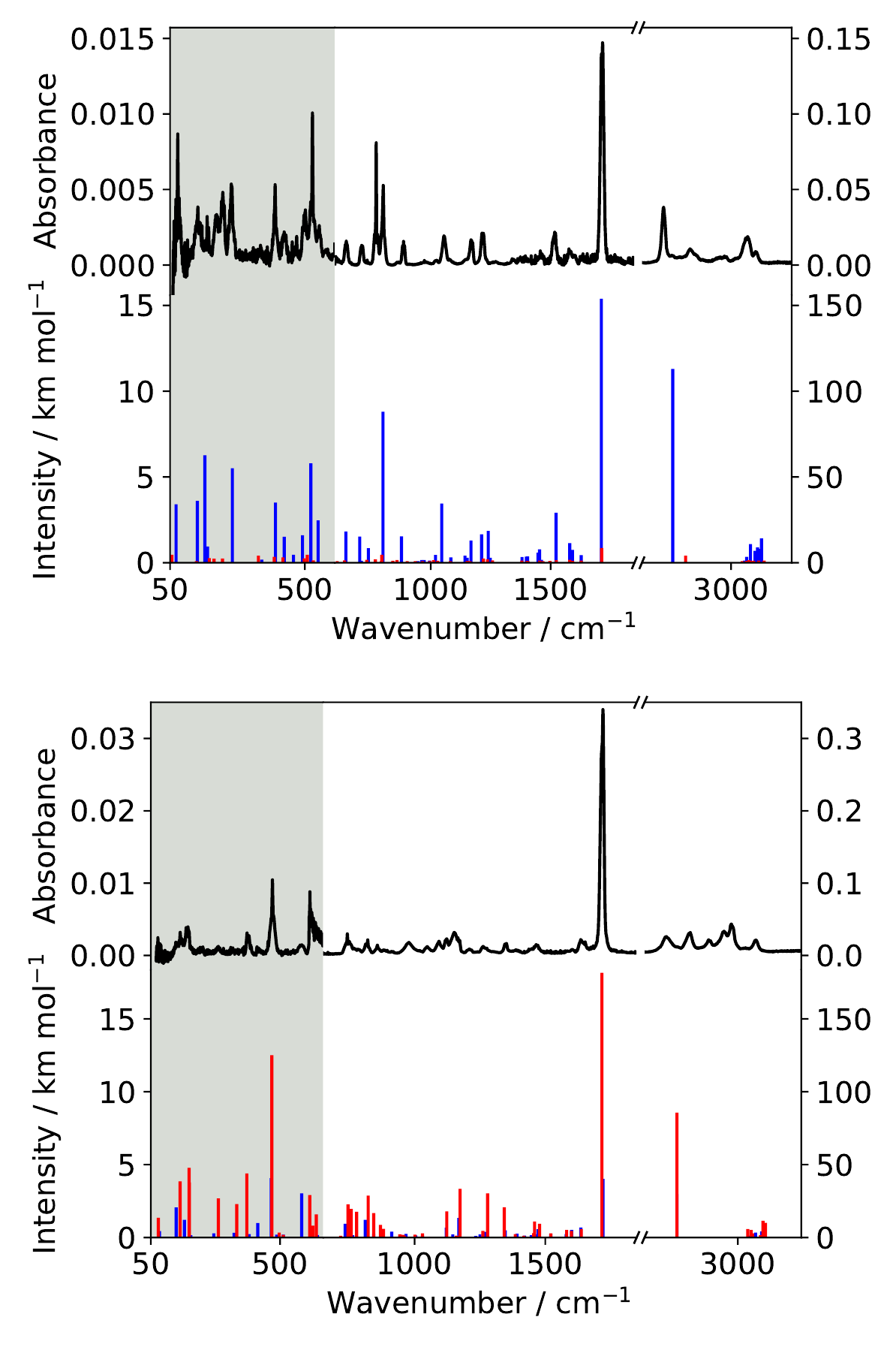}
\caption{Experimental infrared spectra (in black) of 1- (top plot) and 2-naphthaldehyde (bottom plot), and comparison with a stick spectrum of the predicted fundamental bands (hybrid level for positions and MP2/VTZ for harmonic intensities) of the \textit{cis} (in blue) and \textit{trans} (in red) conformers of each species. For the sake of clarity, all intensities are multiplied by a factor 10 in the low frequency part of the spectrum (gray background areas). Intensity ratios between isomers of each species are calculated from the estimated population at room temperature. 
}
	\label{fig:IR_Naphtaldehydes}
\end{figure}

Using the experimental conditions described in the previous section, the vibrational bands observed for both 1- and 2-naphthaldehyde samples come out as unresolved rotational envelopes (spreading over about \SI{20}{\wn}, Figure~\ref{fig:IR_Naphtaldehydes}). Most of the IR bands are relatively weak and many possible overlapping bands complicate the vibrational assignment, in particular in the mid-IR. The experimental band frequencies are obtained by measuring the maximum absorption of the relatively sharp Q-branch for the out-of-plane modes ($c$-type bands). For the in-plane modes ($a/b$-type bands), we used the Q-branch when the latter was clearly observed but, in many cases these bands appear as broad, structureless envelopes in which cases the band frequency was measured at the maximum of the band. The resulting frequency accuracy ranges from about \SI{1}{\wn} for well-defined Q-branches to several \si{\wn} otherwise.

Only the most stable conformers of each compound, $cis$-1-naphthaldehyde and $trans$-2-naphthaldehyde, whose IR bands dominate the experimental spectra, have been detected using our room temperature experimental setup (Figure~\ref{fig:IR_Naphtaldehydes}).
Table~\ref{tab:NaphthalVibFreq} reports the proposed vibrational assignments and the comparison with the calculations of the far-IR bands (below \SI{700}{\wn}) while the full IR assignment is reported in Table~S4 in the Supporting Information. The comparison between experimental and calculated IR spectra of 1- and 2-naphthaldehyde species are displayed on Figure~\ref{fig:IR_Naphtaldehydes}. 

\begin{table*}
  \centering
    \caption{Calculated Far-IR Fundamental Vibrational Modes of $cis$-1-Naphthaldehyde and $trans$-2-Naphthaldehyde and Comparison with Proposed Experimental Assignments. Energies are given in \si{\wn} and intensities in \si{\km\per\mol}.}
    \begin{tabular}{lc rrcrd{2.1}d{3.0}r c rrcrd{2.1}d{3.0}r}
    \toprule
        &  & \multicolumn{7}{c}{$cis$-1-naphthaldehyde} && \multicolumn{7}{c}{$trans$-2-naphthaldehyde}\\ \cmidrule{3-9} \cmidrule{11-17}
        &  & \multicolumn{2}{c}{B97-1}  && \multicolumn{2}{c}{MP2} & \multicolumn{1}{c}{Hybrid}  &  \multicolumn{1}{c}{Exp.} && \multicolumn{2}{c}{B97-1}  && \multicolumn{2}{c}{MP2} & \multicolumn{1}{c}{Hybrid}  &  \multicolumn{1}{c}{Exp.} \\ \cmidrule{3-4}\cmidrule{6-7} \cmidrule{11-12} \cmidrule{14-15}
    $\nu$ & sym. & \multicolumn{1}{c}{$E_\mathrm{h}$} & \multicolumn{1}{c}{$E_\mathrm{a}$} && \multicolumn{1}{c}{$E_\mathrm{h}$} & \multicolumn{1}{c}{$I_\mathrm{h}$} & \multicolumn{1}{c}{$E$} & \multicolumn{1}{c}{$E$}& & \multicolumn{1}{c}{$E_\mathrm{h}$} & \multicolumn{1}{c}{$E_\mathrm{a}$} && \multicolumn{1}{c}{$E_\mathrm{h}$} & \multicolumn{1}{c}{$I_\mathrm{h}$} & \multicolumn{1}{c}{$E$} & \multicolumn{1}{c}{$E$}     \\
    \midrule
    32& A'  & 658 & 649 && 655 & 18.3 & 646 & 647  &&  639 & 633 && 632 & 1.9  & 625 & \\
    33& A'  & 553 & 546 && 550 & 2.5  & 544 &  548 &&  612 & 606 && 608 & 3.6  & 602 & 603\\
    34& A'  & 501 & 495 && 498 & 1.6  & 491 &  497 &&  519 & 514 && 516 & 0.1  & 510 & \\
    35& A'  & 436 & 433 && 434 & 1.5  & 431 &  430 &&  390 & 388 && 385 & 5.5  & 383 & 388\\
    36& A'  & 366 & 362 && 359 & 0.1  & 355 &      &&  348 & 350 && 347 & 2.8  & 348 & \\
    37& A'  & 218 & 165 && 218 & 6.5  & 166 & 225 ?&&  176 & 184 && 175 & 6.0  & 182 & 177\\
    \\
    47& A'' & 639 & 637 && 613 & 0.01 & 611 &      &&  645 & 639 && 619 & 0.9  & 614 & \\
    48& A'' & 538 & 529 && 528 & 6.0  & 519 & 525  &&  512 & 507 && 501 & 0.3  & 496 & \\
    49& A'' & 479 & 472 && 468 & 0.4  & 462 &      &&  485 & 479 && 476 & 15.9 & 471 &  472   \\
    50& A'' & 413 & 408 && 406 & 3.6  & 401 & 400  &&  400 & 395 && 392 & 0.04 & 387 & \\
    51& A'' & 273 & 266 && 265 & 5.7  & 258 & 255  &&  297 & 289 && 293 & 3.3  & 285 & \\
    52& A'' & 181 & 177 && 178 & 0.9  & 174 &      &&  191 & 187 && 187 & 4.7  & 183 & 181   \\
    53& A'' & 151 & 146 && 147 & 3.7  & 141 & 143  &&  160 & 156 && 156 & 4.8  & 152 & 152   \\
    54& A'' & 85  & 76  && 79  & 3.5  & 70  & 76   &&  78  & 77  && 77  & 1.6  & 75  & \\          
    \bottomrule
    \end{tabular}%
    
    \medskip
    \begin{minipage}{0.9 \textwidth}
    \textbf{Note.} $\mathrm{h}$ and $\mathrm{a}$ subscripts stand for harmonic and anharmonic, respectively.
    \end{minipage}
      \label{tab:NaphthalVibFreq}%
\end{table*}%

Due to the relatively low vapor pressure of the naphthaldehyde samples at room temperature, our detection was limited to the most intense mid- and far-IR bands (of calculated intensities larger than about 3 km.mol$^{-1}$). In the far-IR, the vibrational bands involve in-plane and out-of-plane deformations of the carbon skeleton of the molecules. Assignments are proposed on the basis of the the agreement with the calculated frequencies (hybrid method), the relative harmonic intensities of the modes (MP2 calculation), and the rotational contour of the band (since out-of-plane modes are $c$-type and thus display distinctive sharp and intense Q-branches). It is worth noting that for most of the calculated vibrational modes, the anharmonic corrections are relatively small (typically of order of a few \si{\wn}), at the notable exception of some bands (see, $e.g.$, in Table \ref{tab:NaphthalVibFreq} the comparison between harmonic and anharmonic energies of $v_{37}=1$ of $cis$-1-naphthaldehyde molecule for which the anharmonic shift is larger than \SI{50}{\wn} making the assignment rather insecure). For most of the bands, the agreement between observed and calculated bands (both in terms of frequency and intensity) is quite good (corresponding to a few wavenumbers difference) supporting the use of such hybrid calculations (as already shown in previous papers, see $e.g.$, ref~\citenum{Gruet16} and references therein) to predict relatively accurately IR spectra of large and rigid molecules at a low computational cost. 

\subsection{Pure Rotation Spectra}

The analysis procedure used in this work is similar to the one used in previous investigations of rotationally resolved spectra of N-substituted PAHs~\cite{Gruet16}. Briefly, the theoretical calculations provide a first set of rotational constants enabling the search for the centimeter-wave jet-cooled rotational transitions of the molecules. Once the most intense lines are observed and assigned, an accurate set of parameters can be derived that then allows a relatively straightforward assignment of the millimeter-wave room-temperature spectra using a combination of Loomis-Wood~\cite{Lodyga07}, PGOPHER~\cite{Western17}, and Pickett's suite of programs~\cite{Pickett91}. 

The ground-state conformer of each of the studied compounds is a nearly prolate asymmetric top belonging to the $C_\mathrm{s}$ symmetry point group with a permanent dipole moment along both the $a$ and $b$ axis (see Tables~\ref{tab:RotationNaphthaldehydes} and\ref{tab:RotationNaphtols} for dipole-moment values). The rotational transitions were fitted to a Watson Hamiltonian in the $A$-reduction and I$^r$ representation. 
For the four most stable species studied here, both $a$-type and $b$-type transitions were observed in the FTMW spectra while the lower spectral resolution of the room temperature millimeter-wave spectra, together with the limited SNR, did not allow to resolve these different components.
Typically, transitions involving quantum numbers up to $J \sim 150$ and $K_a \sim 40$ were assigned. Owing to the relatively high rigidity of the species, only the three rotational constants ($A, B, C$)  and the quartic centrifugal distortion terms were required to fit the data to their experimental accuracy. Tables \ref{tab:RotationNaphthaldehydes} and \ref{tab:RotationNaphtols} contain the results of our analysis and the comparison with the hybrid calculations. The full tables of rotational constants obtained at each stage of the calculations are provided in Table~S5 in the Supporting Information. The GS rotational constants calculated using this hybrid method are in very good agreement with the experimental values, with errors below 0.5\,\% for the $A$, $B$, and $C$ rotational parameters. However, as such rigid molecules have relatively small quartic centrifugal distortion constants, the agreement between the measured and calculated parameters is significantly worse (in particular the $\delta_J$ and $\delta_K$ parameters can show opposite signs).

For all species, only the conformer of lowest energy was observed unambiguously in the experimental spectra. In the millimeter-wave region, where broad surveys were recorded, the lack of detection of the high-energy conformer is imputed to the low vapor pressure at room temperature, which results in relatively limited SNR, despite the relatively large value of the permanent dipole moments of each species. 
In the case of the millimeter-wave spectrum of $trans$-1-hydroxynaphthalene, the SNR is somewhat larger than for the three other species, allowing pure rotational lines involving four different ES to be assigned. Unambiguous assignments of these ES pure rotation spectra are allowed thanks to the good predicting power of the calculated constants by comparison with anharmonic calculations (Figure~\ref{fig:mmw1Naphtol} and Tables~S6 and~S7 in the Supporting Information).



\begin{table*}
  \centering
  \caption{Experimental and Calculated Rotational Parameters (in \si{\MHz}) for the Naphthaldehyde Species. Error ($1\sigma$) on parameters are given between parenthesis in units of the last digit.}
    \begin{tabular}{l d{4.3}d{4.10}d{3.2} r d{4.3}d{4.10}d{3.2}}
    \toprule
          & \multicolumn{3}{c}{1-$cis$-naphthaldehyde} & & \multicolumn{3}{c}{2-$trans$-naphthaldehyde} \\ \cmidrule{2-4} \cmidrule{6-8}
          & \multicolumn{1}{c}{Calc.$^a$} &    \multicolumn{1}{c}{Exp. }&    \multicolumn{1}{c}{$\delta^b$ / \%} && \multicolumn{1}{c}{Calc.$^a$}& \multicolumn{1}{c}{Exp.}&    \multicolumn{1}{c}{$\delta^b$  / \%} \\ \midrule
    $A_0$                       & 1380     &   1384.40872\,(21) & 0.32 && 2802     & 2810.43194\,(15)  & 0.30\\
    $B_0$                       &  995     &   999.015243\,(71) & 0.40 &&  580     & 581.095981\,(24)  & 0.19\\
    $C_0$                       &  579     &   580.551953\,(19) & 0.27 &&  480     & 481.700564\,(18)  & 0.35\\
    $\Delta_J    \times 10^{6}$ &  0.46    &    23.855\,(38)    & 5086 && 2.3      &     4.44535\,(45) & 93  \\
    $\Delta_{JK} \times 10^{3}$ & -0.11    &   -0.0482\,(13)    & -56  && 0.14     &    0.025237\,(11) & -82 \\
    $\Delta_K    \times 10^{3}$ & 0.065    &    0.0882\,(35)    & 36   && 1.3      &    0.191170\,(84) & -85 \\
    $\delta_J    \times 10^{6}$ & -0.52    &     9.751\,(19)    & -1975&& 0.21     &     0.81290\,(19) & 287 \\
    $\delta_K    \times 10^{3}$ & -0.050   &   0.01153\,(72)    & -123 && -0.035    &    0.025091\,(24) & -172\\
    \\
    dipole \\
    \hspace{3mm}$a$-type / D  &   \multicolumn{1}{c}{1.3}  && &       &    \multicolumn{1}{c}{3.5} &\\
    \hspace{3mm}$b$-type / D  &   \multicolumn{1}{c}{2.7}  && &       &    \multicolumn{1}{c}{0.7} &\\
    \# lines\\
    \hspace{3mm}FTMW & &\multicolumn{1}{c}{62} &&     &        & \multicolumn{1}{c}{32}   \\
    \hspace{3mm}mmW  & &\multicolumn{1}{c}{3 018} &&       &&        \multicolumn{1}{c}{2671}    \\
    RMS / kHz && \multicolumn{1}{c}{62}    &&              && \multicolumn{1}{c}{69}      \\
    \bottomrule
    \end{tabular}%
    
    \medskip
    \begin{minipage}{0.8 \textwidth}
    $^a$ Hybrid rotational constants and MP2 centrifugal distorsion terms, see text\\
    $^b \delta = (B_{\mathrm{Exp.}}-B_{\mathrm{Calc}}) / B_{\mathrm{Calc.}} \times 100$
    \end{minipage}
  \label{tab:RotationNaphthaldehydes}%
\end{table*}%

\begin{table*}
  \centering
  \caption{Experimental and calculated rotational parameters (in MHz) for the hydroxynaphthalene species. Error ($1\sigma$) on parameters are given between parenthesis in units of the last digit. }
    \begin{tabular}{l d{4.3}d{4.10}d{3.2} r d{4.3}d{4.10}d{3.2}}
    \toprule
          & \multicolumn{3}{c}{1-$trans$-hydroxynaphthalene} & & \multicolumn{3}{c}{2-$cis$-hydroxynaphthalene} \\ \cmidrule{2-4} \cmidrule{6-8}
          & \multicolumn{1}{c}{Calc.$^a$} &    \multicolumn{1}{c}{Exp. }&    \multicolumn{1}{c}{$\delta^b$ / \%} && \multicolumn{1}{c}{Calc.$^a$}& \multicolumn{1}{c}{Exp.}&    \multicolumn{1}{c}{$\delta^b$  / \%} \\ \midrule
    $A_0$                       & 1934     &  1942.100623\,(53) & 0.42  && 2837     & 2849.155543\,(64) & 0.43\\
    $B_0$                       & 1132     &  1133.623211\,(21) & 0.14  && 823      &  824.632161\,(25) & 0.20\\
    $C_0$                       & 714      &   716.017762\,(14) & 0.28  && 638      &  639.723697\,(15) & 0.27\\
    $\Delta_J    \times 10^{6}$ & 18       &     17.90338\,(92) & -0.54 && 6.6      &      8.2474\,(19) & 25  \\
    $\Delta_{JK} \times 10^{3}$ & 0.010    &    0.0159910\,(44) & 60    && 0.071    &    0.014844\,(25) & -79 \\
    $\Delta_K    \times 10^{3}$ & 0.050    &     0.044395\,(14) & -11   && 0.22     &    0.185375\,(72) & -16 \\
    $\delta_J    \times 10^{6}$ & 6.6      &      6.43578\,(37) & -2.5  && 1.3      &     2.05543\,(90) & 58  \\
    $\delta_K    \times 10^{3}$ & 0.034    &    0.0351740\,(62) & 3.5   && 0.040    &    0.030525\,(43) & -24 \\
    \\
    dipole \\
    \hspace{3mm}$a$-type / D  &   \multicolumn{1}{c}{1.2}  && &       &    \multicolumn{1}{c}{0.2} &\\
    \hspace{3mm}$b$-type / D  &   \multicolumn{1}{c}{0.2}  && &       &    \multicolumn{1}{c}{1.0} &\\
    \# lines\\
    \hspace{3mm}FTMW & &\multicolumn{1}{c}{50} &&     &        & \multicolumn{1}{c}{82}   \\
    \hspace{3mm}mmW  & &\multicolumn{1}{c}{3 636} &&       &&        \multicolumn{1}{c}{1248}    \\
    RMS / kHz && \multicolumn{1}{c}{62}    &&              && \multicolumn{1}{c}{37}      \\
    \bottomrule
    \end{tabular}%
    
    \medskip
    \begin{minipage}{0.8 \textwidth}
    $^a$ Hybrid rotational constants and MP2 equilibrium centrifugal distortion terms, see text\\
    $^b \delta = (B_{\mathrm{Exp.}}-B_{\mathrm{Calc}}) / B_{\mathrm{Calc.}} \times 100$
    \end{minipage}
  \label{tab:RotationNaphtols}%
\end{table*}%



\section{Discussion}

The relative stability of the species in the naphthaldehyde / hydroxynaphthalene families warrants further discussion. Calculated energies performed in this work show that the substitution in position 2- is the most stable for the \ce{-COH} isomers while the \ce{-OH} are almost isoenergetic, and the $cis$ vs. $trans$ conformer stability is completely different for both the two considered substitutions (\ce{-COH} and \ce{-OH}) and their isomers (1- and 2-, Figure~\ref{fig:ConformLand}). The latter point is unambiguously confirmed by the experimental measurements since for all pure compounds studied, only the conformer predicted to be the most stable by the calculations has been observed.  In order to shed some light into the physical and chemical factors that govern these conformational landscapes, in this section we discuss the differences between all species in terms of electrostatic interactions (steric repulsion and local charges) that we found in agreement with the relative stabilities.

\begin{figure*}[htp]
\includegraphics[width=0.85\textwidth]{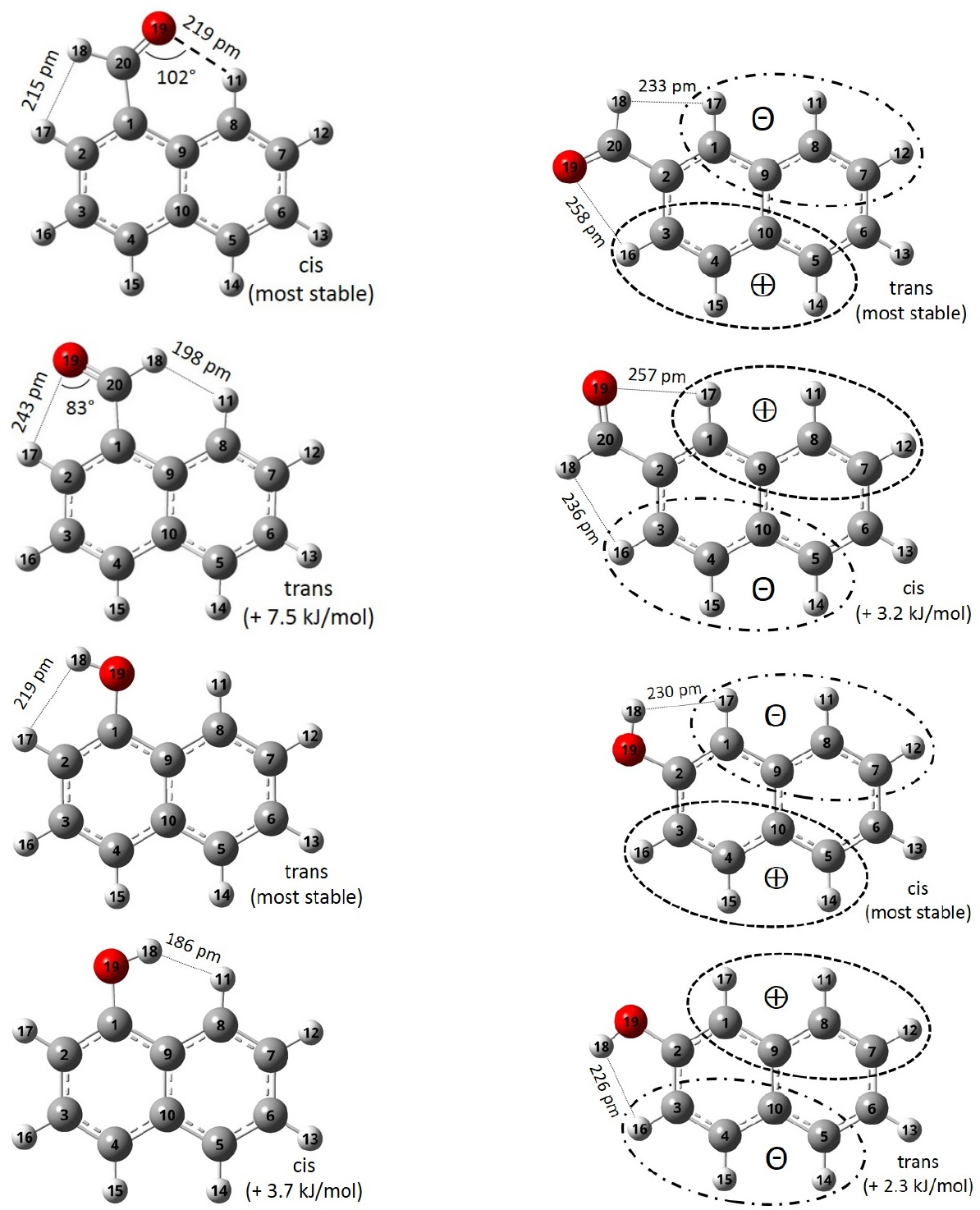}
	\caption{Calculated structures at equilibrium (MP2/aug-cc-pVQZ) of the two conformers of 1-naphthaldehyde, 2-naphthaldehyde, 1-hydroxynaphthalene, and 2-hydroxynaphthalene. Atom numbering, relative energy (hybrid level), schematic representation of the charges distribution, and relevant topological parameters are indicated.}
	\label{fig:conf-all}
\end{figure*}

\subsection{Isomeric Considerations}
Prior to a detailed conformational analysis, only the most stable conformer of each species will be considered here. The charges derived from the NBO calculations (see Tables~S8 and~S9 in the Supporting Information) together with molecular structures (see Figure~\ref{fig:conf-all} for the most relevant topological parameters) are useful indicators to assess the relative isomeric stability, by verifying their agreement with relative energies and well-established chemical considerations: (i) an aldehyde is an electrophilic group; (ii) an hydroxy group can either be electron sharing or withdrawing (by resonant or inductive effect, respectively); and (iii) a hydroxy is an activating group while an aldehyde is a deactivating group. On one hand, steric effects within isomers might be estimated from the electrostatic interaction (Coulomb type) between the H atom $i$ of the functional group and the closest H atom $j$ attached to the ring (repulsive): $E_{q} = \frac{q_{i} \times q_{j}}{r_{ij}}$ where $q$ are the calculated charges and $r_{ij}$ is the calculated distance between charges centers. Then, the difference of these electrostatic energies between isomers $\Delta E_q$ might be used to account for the destabilization of one compared to the other. On the other hand, atomic charges of bare naphthalene (calculated using the same method: NBO6 at the MP2/aug-cc-pVQZ level) can be used to point out at which position (1- or 2-) the attack of a functional group might be favored. Atomic charges are gathered in Tables~S8 and~S9 in the Supporting Information; atom labels and relevant distances are shown in Figure~\ref{fig:conf-all}.

\paragraph{Naphthaldehydes.} The 2-naphthaldehyde isomer is calculated to be the most stable form, by far, with an energy difference of more than \SI{6}{\kJ\per\mol} with the 1-naphthaldehyde species (Figure~\ref{fig:ConformLand}). This destabilization of 1- over 2- qualitatively agrees with a steric repulsion larger in 1- than in 2-: $\Delta E_{q} = E_{q}(cis-1-) - E_{q}(trans-2-) =$ \SI{2.3}{\kJ\per\mol}. This isomeric preference is also in line with consideration (i) when looking at the charges distribution in naphthalene (see values in Table~S8 in the Supporting Information): the charge of the \ce{C-H} group at position 2- (C2+H17) is negative (\SI{-0.0105}{\astronomicalunit}) while it is positive (\SI{+ 0.0201}{\astronomicalunit}) at position 1- (C1+H18) . Thus, the electrophilic attack of an aldehyde group on naphthalene would be favored in position 2-. Therefore, both energies and charges agree with consideration (i) to designate isomer 2- as the favored form.

\paragraph{Hydroxynaphthalenes.} Here again, isomer 2- is slightly more stable than isomer 1- (\SI{0.1}{\kJ\per\mol} apart at the hybrid level of theory, see  Figure~\ref{fig:ConformLand}). Steric repulsion is also qualitatively pointing toward a destabilization of 1- over 2- with $\Delta E_{q} = E_{q}(trans-1-) - E_{q}(cis-2-) =$ \SI{2.9}{\kJ\per\mol} . In this case, the sign of charges at positions 1- and 2- in naphthalene is probably a less efficient indicator since consideration (ii) says that an attack of a hydroxy group can either be electron sharing or withdrawing, so it can occur either at position 1- or 2- regardless the sign of the charges. However, it may corroborate that the two hydroxynaphthalene isomers are almost iso-energetic.

\begin{figure}[htbp]
    \includegraphics[width=\columnwidth]{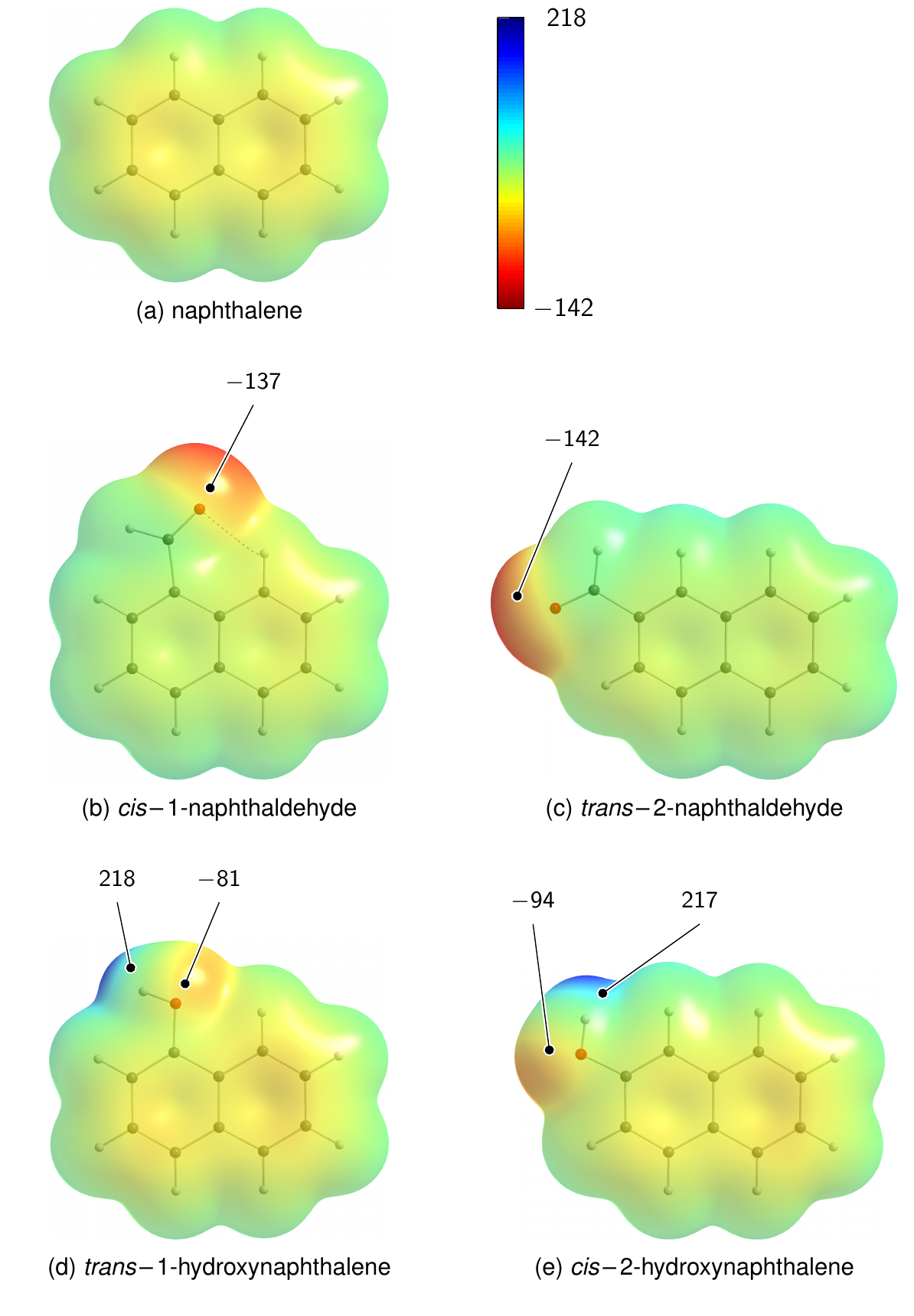} 
    \caption{Molecular electrostatic potentials (MEP) mapped on the \SI{0.001}{\hartree} isodensity surface calculated at the MP2/aug-cc-pVQZ level of theory for naphthalene and the most stable conformers of the hydroxynaphthalene and naphthaldehyde.
    MEP values range from \SI{-142}{\kJ\per\mol} (red) to \SI{218}{\kJ\per\mol} (blue).} 
    \label{Fig:MEPmaps}
\end{figure}

\paragraph{Both Molecules.} The large difference in electronic structure among naphthaldehyde, hydroxynaphthalene, and bare naphthalene prevents the use of relative energies and local charges differences to assess isomeric preference. However,  the global charge redistribution upon functionalization might be a useful qualitative indicator to assess the validity of NBO calculations. Figure~\ref{Fig:MEPmaps} displays the molecular electrostatic potentials (MEP) mapped on the \SI{0.001}{\hartree} isodensity surface (MP2/aug-cc-pVQZ) for naphthalene and the most stable conformers of hydroxynaphthalene and naphthaldehyde, created by the AIMall visualization software~\cite{prog-AIMAll}. Looking at the aromatic rings, MEP for hydroxynaphthalene species stay roughly the same as for naphthalene while they get globally less negative for naphthaldehyde with most of the negative charges concentrating themselves on the oxygen atom of the aldehyde group. This tends to indicate that an electrophilic attack would be less efficient on naphthaldehyde than on hydroxynaphthalene or bare naphthalene, which qualitatively agrees with consideration (iii).

\subsection{Conformational Landscape}
Once confidence is gained in the molecular orbitals and charges derived from NBO calculations, they can be used to discuss the conformational landscape of the four compounds. Indeed, the way that the charges are distributed in the aromatic ring upon functionalization together with estimations of electrostatic interactions can be used for understanding why for each isomer one conformer is more stable than the other. As shown in the following, the situation for isomers 1- is relatively straightforward, while isomers 2- situation calls for further discussion.

\begin{figure}[htbp]
	\includegraphics[width=\columnwidth]{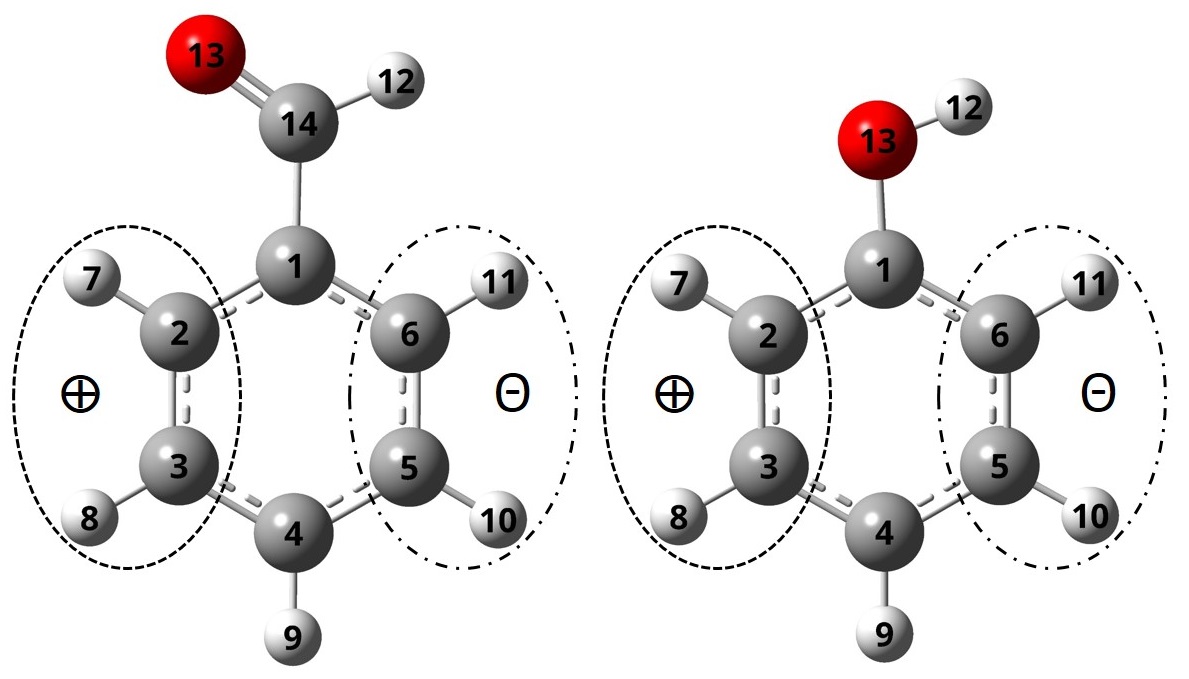} 
	\caption{Schematic representation of the charges distribution in benzaldehyde (left panel) and hydroxybenzene (right panel) as reported in Table \ref{tab:qbenz}.}
	\label{fig:benzox}
\end{figure}

\paragraph{1-Naphthaldehyde.} An intramolecular hydrogen bond between the oxygen atom of the aldehyde group and the H11 hydrogen atom attached to the C8 carbon (Figure~\ref{fig:conf-all}) stabilizes the \textit{cis} conformer compared to the \textit{trans}. The existence of an intramolecular hydrogen bond in the \textit{cis} isomer is evidenced by NBO calculations showing a charge transfer from the O atom lone pair (donor) to the C8-H11 antibonding orbital (acceptor), corresponding to a stabilization energy E(2) of about \SI{5.1}{\kJ\per\mol}. This value corresponds to a typically weak intramolecular hydrogen bond, as reflected as well by the long hydrogen bond length, \SI{219}{\pm} (see Figure~\ref{fig:conf-all}) and its orientation~\cite{Grabowski2005}. However, in the \textit{trans} structure, the distance is too long and the angle is too obtuse, preventing the detection of a significant orbital mixing in the NBO analysis (E(2) is smaller than the \SI{2}{\kJ\per\mol} threshold)~\cite{nbo-Reed-CR1988-88-899}. Qualitatively, the hydrogen bond stabilization energy of \SI{5.1}{\kJ\per\mol} accounts for most of the $cis$/$trans$ relative energy (\SI{7.5}{\kJ\per\mol}). The difference (\SI{2.4}{\kJ\per\mol}) might be explained by adding an extra destabilization of \textit{trans} over \textit{cis} due to a larger a steric repulsion at the H atom side in \textit{trans} than in \textit{cis}: $\Delta E_{q} = E_{q}(trans) - E_{q}(cis) =$ \SI{1.7}{\kJ\per\mol}.

\paragraph{1-Hydroxynaphthalene.} It is clear from topological observations that no intramolecular hydrogen bond is possible with hydroxy as the functional group. However, a stronger steric repulsion between the H atom of the hydroxy group and its closest H neighbor destabilizes the \textit{cis} conformer compared to the \textit{trans}. This effect is visible when looking at the \ce{H-H} distance (see Figure~\ref{fig:conf-all}), which equals 186\,pm in the \textit{cis} isomer, a distance that is far below twice the an der Waals radius of an H atom ($2 \times 120 = 240$\,pm), while it is significantly longer in the \textit{trans} isomer (219\,pm). Here, the destabilization by electrostatic interaction with $\Delta E_{q} = E_{q}(cis) - E_{q}(trans) =$ \SI{2.2}{\kJ\per\mol}accounts for most of the $cis$/$trans$ relative energy (\SI{3.7}{\kJ\per\mol}).

\begin{table}[htp]
  \centering 
  \caption{Calculated Sum of Charges around the Functional Group in Functionalized Benzene (see Tables~S8 and~S9 in the Supporting Information for atomic charges and see Figure~\ref{fig:benzox} for atoms labelling). Values are given in atomic units.}
    \begin{tabular}{l cc}
    \toprule
    atoms	&	benzaldehyde	&	hydroxybenzene	\\ \midrule
    \multicolumn{3}{c}{``closest neighbors''} \\
    H12 side$^a$	&	0.000	&	-0.103	\\
    opposite$^b$	&	0.025	&	-0.060	\\
    $\Delta$q$^c$	&	\textbf{-0.025}	&	\textbf{-0.043}	\\
    \\
    \multicolumn{3}{c}{``wider view''} \\
    H12 side$^d$	&	0.003	&	-0.079	\\
    opposite$^e$	&	0.036	&	-0.035	\\
    $\Delta$q$^c$	&	\textbf{-0.033}	&	\textbf{-0.044}	\\
    \bottomrule
    \end{tabular}
    \medskip
    
    \vspace{1mm}
    \begin{minipage}{0.8 \columnwidth}
    $^a$ C6+H11\\
    $^b$ C2+H7\\
    $^c$ $\Delta$q = (H12 side - opposite)\\
    $^d$ C6+H11+C5+H10\\
    $^e$ C2+H7+C3+H8\\
    \end{minipage}
    \label{tab:qbenz}%
\end{table}%

\paragraph{Isomers 2-. }
A closer look at charges seems necessary to bring some understanding in these conformational landscapes. In addition to simple electrostatic interactions between local charges, longer range effects might be expected because of the aromatic system of such species. Contrary to isomers 1- for which geometrical asymmetry around the functional groups induces significant differences between conformers (stabilization/destabilization due to attractive/repulsive interactions), isomers 2- with conformers having quasi-equal topological parameters appears as ideal systems to eventually study the link between charges distribution and conformational preferences.

The more simple functionalized benzenes (\textit{i.e.}, benzaldehyde and hydroxybenzene; see Tables~S8 and~S9 in the Supporting Information) may be used as reference systems. In the case of benzene, owing to its higher symmetry, a 180\textdegree\ rotation of the functional group gives rise to two iso-energetic mirror images, \textit{i.e.} functionalized benzenes have no conformational forms. Then, the charge distribution within the aromatic ring (uniform in bare benzene) perturbed by the functional group can be evaluated and may serve as reference. In both functional groups, the H atom is a carrier of positive charges (H12; see Tables~S8 and~S9). The sum of charges on the closest neighbor \ce{C-H} groups in the ring (see Table \ref{tab:qbenz}) is less positive (or more negative) at the side of H12 (C6+H11) than at its opposite side (C2+H7). However, electrons delocalization associated with aromaticity should facilitate charge transfer between members of the ring. Thus, it appears relevant to take this effect into account by looking at a ``wider view'', \textit{i.e.}, by adding charges of the next \ce{C-H} group to those of the closest neighbor (see Figure~\ref{fig:benzox}). The same results are observed in this ``wider view'', as shown in Table \ref{tab:qbenz}. Such asymmetric charge distribution within the aromatic ring from one side of the functional group to the other is most probably due to the repulsive action of H12 as a positive charge center so that an energetic stabilization of the whole system is achieved by transferring some of the ring positive charge to the opposite side, which is facilitated by aromaticity. Then, such stabilization evidenced by an asymmetry of charges is expected in functionalized naphthalene, the most stable conformer being the one showing the same trend or at least the most pronounced one.
\begin{table}[htp]
  \centering 
  \caption{Calculated Sum of Charges around the Functional Group in Functionalized Naphthalene (see Tables~S8 and~S9 in the Supporting Information for atomic charges and Figure~\ref{fig:conf-all} for atoms labelling). Most stable conformers are specified in bold. Values are given in atomic units.}
    \begin{tabular}{l cc c cc}
    \toprule
    &  \multicolumn{2}{c}{naphthaldehyde} && \multicolumn{2}{c}{hydroxynaphthalene} \\ \cmidrule{2-3} \cmidrule{5-6}
    & \textbf{trans-2} & \textit{cis-2} && \textbf{cis-2} & \textit{trans-2}   	\\ \midrule
    \multicolumn{6}{l}{\bf``closest neighbors''}  \\
    H18 side	&	0.098$^a$	&	-0.005$^b$	&&	-0.130$^a$	&	-0.071$^b$ 	\\
    opposite	&	0.028$^b$	&	0.104$^a$	&&	-0.070$^b$	&	0.000$^a$ 	\\
    $\Delta$q$^c$	&	+0.070	&	-0.111	&&	-0.060	&	-0.071 	\\
    \\
    \multicolumn{6}{l}{\bf ``wider view''}  \\
    H18 side	&	0.027$^d$	&	0.004$^e$	&&	-0.099$^d$	&	-0.027$^e$	    \\
    opposite	&	0.121$^e$	&	0.063$^d$	&&	-0.039$^e$	&	0.003$^d$ \\
    $\Delta$q$^c$	&	\textbf{-0.094}	&	-0.059	&&	\textbf{-0.069}	&	-0.024 	\\
    \bottomrule
    \end{tabular}
        \medskip
        
        \vspace{1mm}
    \begin{minipage}{0.8 \columnwidth}
    $^a$ C1+H17\\
    $^b$ C3+H16\\
    $^c$ $\Delta$q = (H18 side - opposite)\\
    $^d$ C1+H17+C9+C8+H11+C7+H12\\
    $^e$ C3+H16+C4+H15+C10+C5+H14\\
    \end{minipage}
      \label{tab:qnaphth}%
\end{table}%

\paragraph{2-Hydroxynaphthalene.} The \textit{cis} conformer is calculated to be the most stable form with an energy difference of \SI{2.3}{\kJ\per\mol}. It qualitatively  agrees with the electrostatic interaction difference of $\Delta E_{q} = E_{q}(trans) - E_{q}(cis) =$ \SI{2.8}{\kJ\per\mol}. Concerning charges distribution, by looking only at the closest neighbor \ce{C-H} groups, there is a positive charge deficit on the side of the hydroxyl hydrogen atom (H18) for both conformers (see Table \ref{tab:qnaphth}), which fails to correlate with the conformational preference. On the contrary, the asymmetry of charges is slightly more pronounced in the less stable conformer (\textit{trans}). However, in the ``wider view'' including the next groups (see Figure~\ref{fig:conf-all}), the positive charge deficit on the side of H18 is more pronounced for the most stable \textit{cis} conformer: (H18 side - opposite) = \SI{-0.069}{\astronomicalunit} for \textit{cis} against \SI{-0.024}{\astronomicalunit} for \textit{trans}. This could be explained by the peculiar $\pi$ delocalization linked to aromaticity. Compared to a single aromatic ring (benzox species), the charge transfer within the $\pi$ orbitals network might be less efficient in a doubly aromatic system (naphthox species), so that the ``closest neighbors'' might be still strongly influenced by the local charges distribution within the functional group and only a ``wider view'' permits to retrieve the asymmetry of charges. Therefore, the conformational preference, evidenced by energetic considerations, could be correlated to an asymmetry of charges within the aromatic systems only on the basis of this hypothesis.

\paragraph{2-Naphthaldehyde.} The \textit{trans} conformer is calculated to be the most stable form with an energy difference of \SI{3.2}{\kJ\per\mol}. It qualitatively agrees with the electrostatic interaction difference of $\Delta E_{q} = E_{q}(trans) - E_{q}(cis) =$ \SI{1.8}{\kJ\per\mol}. In this case as well, the charge distribution at the closest neighbor \ce{C-H} groups is pointing to the \textit{cis} conformer as the most stable (see Table \ref{tab:qnaphth}), contrary to energetic considerations. However, here again, the reverse situation is observed in the ``wider view'' in agreement with energies (see Figure~\ref{fig:conf-all}), (\textit{i.e.}, a positive charge deficit at the H18 side is more pronounced for the most stable \textit{trans} conformer), so that the same hypothesis of a stabilization by charge transfer within the whole aromatic systems may apply.

\section{Conclusions}

The spectra of four oxygen-containing naphthalene species -- 1- and 2-hydroxynaphthalene, and 1- and 2-naphthaldehyde -- have been recorded in the microwave, millimeter-wave, and infrared spectral ranges. Pure rotational and vibrational assignments have been supported by high level quantum chemical calculations, using a hybrid method based on MP2 (harmonic) and DFT (anharmonic) calculations. The pure rotation and rotation-vibration spectra recorded using various experimental setups allowed us to unambiguously identify the most stable conformers of each species, in excellent agreement with the calculated results for the energetic relative stabilities.
The isomeric preference evidenced by relative free energies (isomers 2- are more stable than isomers 1-) has been rationalized in terms of charge distribution within the aromatic rings. In turn, confidence has been gained in the molecular orbitals and charges out of NBO calculations. The comparison with similar calculations for simpler structures (functionalized benzene) leads us to propose an explanation for the conformational preferences of each species on the basis of the influence of the functional groups on the charges distribution within the aromatic rings, providing new insights in the physicochemical properties of isomeric and conformational landscapes of these functionalized PAHs compounds. Finally, our study provides all necessary information for an active search of these molecules in atmospheric and astrophysical environments.

\section*{Acknowledgements}
This work was supported by the Programme National ``Physique et Chimie du Milieu Interstellaire'' (PCMI) of CNRS/INSU with INC/INP cofunded by CEA and CNES; by the CaPPA project (Chemical and Physical Properties of the Atmosphere); and by the CPER ClimiBio funded by the French National Research Agency (ANR) through the PIA (Programme d'Investissement d'Avenir), under contract ANR-11-LABX-0005-01, the I-SITE ULNE/ANR-16-IDEX-0004 ULNE, and the Regional Council Hauts-de-France and the European Funds for Regional Economic Development (FEDER). Part of this work is supported by ``Investissements d'Avenir'' LabEx PALM (ANR-10-LABX-0039-PALM).


\section*{Associated content}
Figures S1 and Tables S1--S9 are reported in the Supporting Information, together with the complete lists of observed transitions and fit files.

\begin{center}
TOC GRAPHIC
\end{center}
\hrulefill

\centering
\includegraphics[width=1.0\columnwidth]{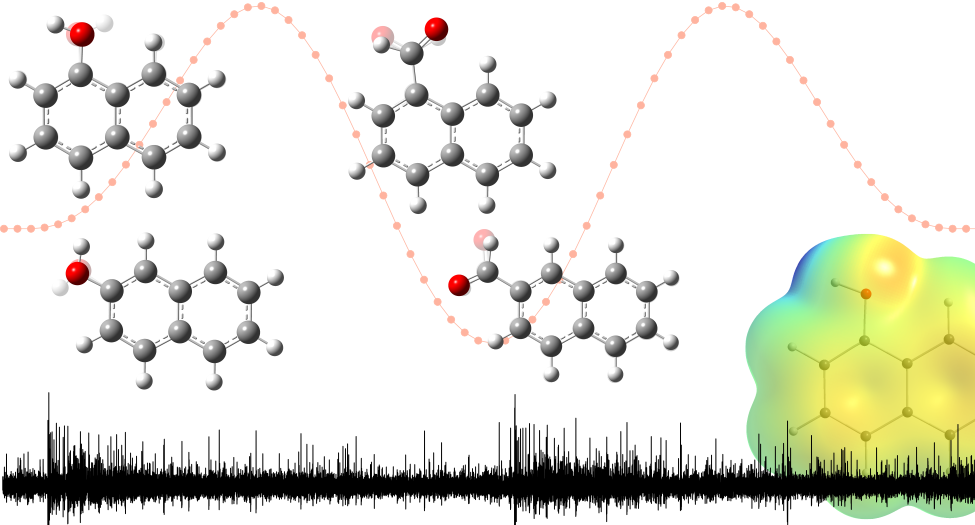}

%
%
%
%
%
%
%

\bibliography{PAHs}


\end{document}


\begin{figure}[ht!]
    \centering
    \includegraphics[width=0.8\textwidth]{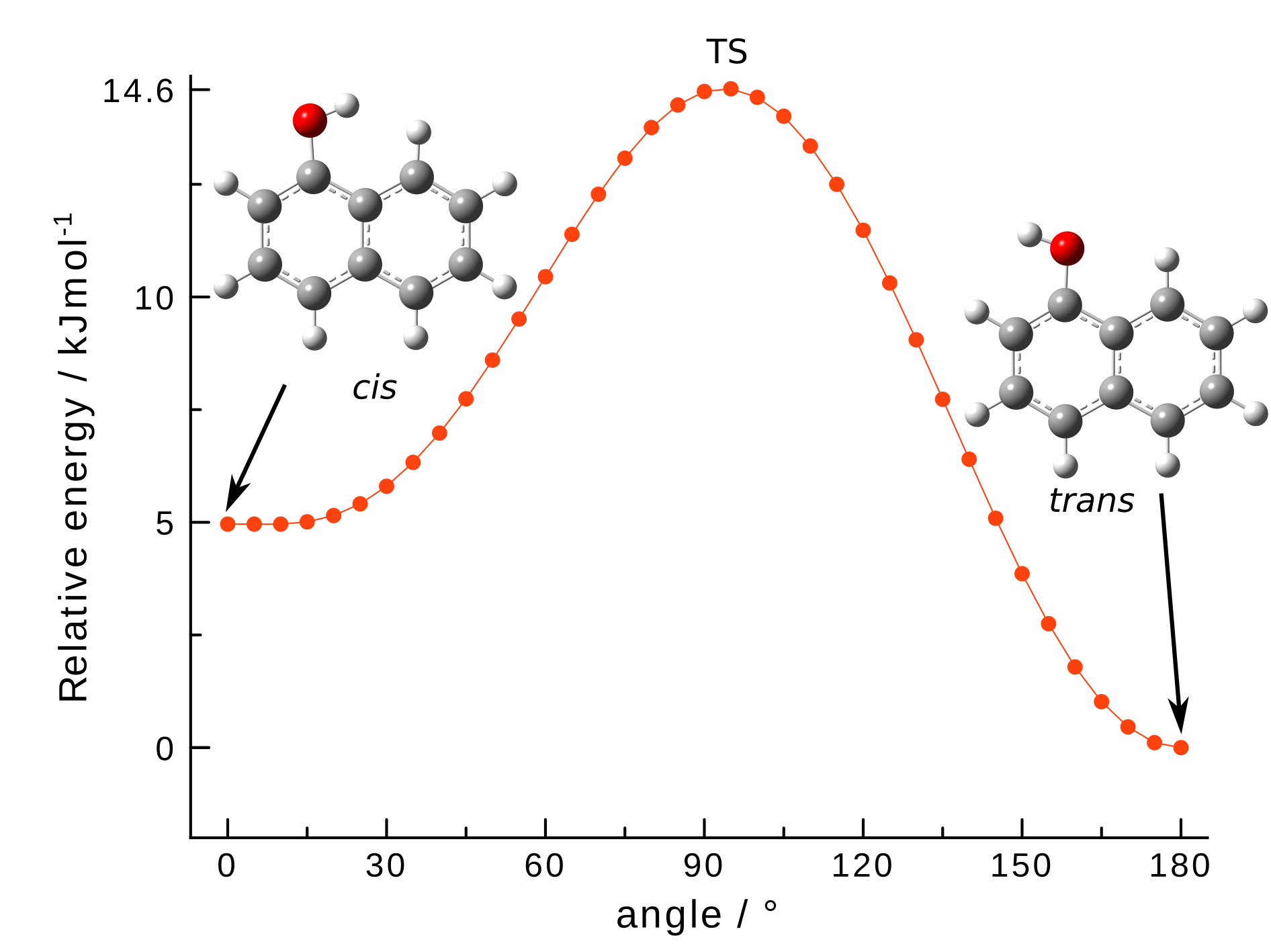}
    \caption[Conformational landscape of 1-hydroxynaphthalene: Potential energy curve resulting from the MP2/cc-pVTZ calculation (values at equilibrium)]{Conformational landscape of 1-hydroxynaphthalene: Potential energy curve resulting from the MP2/cc-pVTZ calculation (values at equilibrium). Energies are indicated relative to $trans$-1-hydroxynaphthalene. Angle values are corresponding to the \ce{C-C-O-H} dihedral angle.}
    \label{fig:landscape}
\end{figure}



\begin{sidewaystable}[ht!]
    \caption[Results from the energy calculations on the hydroxynaphthalene molecules]{Results from the energy calculations on the hydroxynaphthalene molecules. $\Delta_\mathrm{E}$ values are relative to the lower energy conformer of each species. }
    \label{tab:EcalcOL}
    \centering \small
\begin{tabular}{lllrrrcrrr} \toprule
         &                        &          & \multicolumn{3}{c}{\textbf{1-hydroxynaphthalene} }  & & \multicolumn{3}{c}{\textbf{2-hydroxynaphthalene}} \\ \cmidrule{4-6} \cmidrule{8-10}                
         &                        &          & \multicolumn{1}{c}{$cis$ }                 & \multicolumn{1}{c}{TS}           & \multicolumn{1}{c}{$trans$}     && \multicolumn{1}{c}{$cis$}          & \multicolumn{1}{c}{TS}           & \multicolumn{1}{c}{$trans$}    \\    \cmidrule{4-6} \cmidrule{8-10}
         &                        &          &                      &              &                \\
B97-1$^a$& $E_\mathrm{eq}$        & / $E_h$  & -461.1344478 & -461.1302821         & -461.1364003 && -461.1368716 & -461.1301344 & -461.1359809  \\
         & $E_\mathrm{ZPE}$       & / $E_h$  & -460.9839570 & -460.9802350         & -460.9854290 && -460.9860680 & -460.9803100 & -460.9853160  \\
         & $\omega$               & /\si{\wn}& 87.70        & 330.20               & 140.10       && 122.10       & 350.90       & 122.30        \\
         & $\Delta E_\mathrm{eq}$ & / \si{\kJ\per\mol} & 5.1          & 16.1                 &  0           && 0            & 17.69        & 2.34          \\
         &                        & /\si{\wn}& 429          & 1,343                &      0       && 0            & 1,478.64     & 195.49        \\
         & $\Delta E_\mathrm{ZPE}$& / \si{\kJ\per\mol} & 3.9          & 11.7$^d$                 &  0           && 0            & 13.0$^d$         & 1.97         \\
         &                        & /\si{\wn}& 323          & 975$^d$                   &     0        && 0            & 1,088$^d$        & 165.04        \\ \midrule
MP2$^b$  & $E_\mathrm{eq}$        & / $E_h$  & -460.1680434 & -460.1643641         & -460.1699345 && -460.1698188 & -460.1636588 & -460.1688217  \\
         & $E_\mathrm{ZPE}$       & / $E_h$  & -460.0171110 & -460.0135300         & -460.0183590 && -460.0184030 & -460.0130740 & -460.0175580  \\
         & $\omega$               & /\si{\wn}& -57.20       & 332.10               & 138.40       && 120.90       & 340.90       & 120.90        \\
         & $\Delta E_\mathrm{eq}$ & / \si{\kJ\per\mol} & 5.0          & 14.6                 &  0           && 0            & 16.2         & 2.6           \\
         &                        & /\si{\wn}& 415          & 1,223                &    0         && 0            & 1,352        & 219           \\
         & $\Delta E_\mathrm{ZPE}$& / \si{\kJ\per\mol} & 3.3          & 10.7$^d$                 &  0           && 0            & 11.9$^d$         & 2.2           \\
         &                        & /\si{\wn}& 274          & 895$^d$                  &    0         && 0            & 994$^d$          & 185           \\ \midrule
hybrid$^c$& $E_\mathrm{ZPE}$       & / $E_h$  & -460.0175526 & -460.0143170         & -460.0189632 && -460.0190152 & -460.0138344 & -460.0181568  \\
         & $\Delta E_\mathrm{ZPE}$& / \si{\kJ\per\mol} & 3.7          & 10.2$^d$                 &  0           && 0            & 11.5$^d$         & 2.3           \\
         &                        & /\si{\wn}& 310          & 855$^d$                  &   0          && 0            & 962$^d$          & 188           \\
         & Pop.$^e$             & / \%     & 18           &                      & 82           && 71           &              & 29            \\  
         & $\Delta E_{12}\,^f$& / \si{\kJ\per\mol} &     --    &     --          &    0.1      &&         0    &   --     &      --    \\
         &                        & /\si{\wn}&    --    &        --        &    11        &&      0       &  --       &   --     \\
         \bottomrule
    \end{tabular}
    
    \smallskip
    \begin{minipage}{0.8\textwidth}
    $^a$ B97-1/cc-pVTZ anharmonic calculation \hspace{5mm} 
    $^b$ MP2/cc-pVTZ anharmonic calculation  \hspace{5mm} 
    $^c$ hybrid B97-1/MP2/cc-pVTZ calculation, see text \hspace{5mm} 
    $^d$ For the transition states, $\Delta E_\mathrm{ZPE}$ is calculated according to the formula $\Delta E_\mathrm{ZPE} = E(\mathrm{TS})- E(\mathrm{GS}) - \omega(\mathrm{TS})/2$ \hspace{5mm} 
    $^e$ Population at 300 K \hspace{5mm} 
    $^f$ Energy difference between the lower state conformers of 1- and 2-hydroxynaphthalene
    \end{minipage}
\end{sidewaystable}

\begin{sidewaystable}[ht!]
    \caption[Results from the energy calculations on the naphthaldehyde molecules]{Results from the energy calculations on the naphthaldehyde molecules. $\Delta_\mathrm{E}$ values are relative to the lower energy conformer of each species. }
    \label{tab:EcalcAL}
    \centering \small
\begin{tabular}{lllrrrcrrr} \toprule
         &                        &          & \multicolumn{3}{c}{\textbf{1-naphthaldehyde} }  & & \multicolumn{3}{c}{\textbf{2-naphthaldehyde}} \\ \cmidrule{4-6} \cmidrule{8-10}                
         &                        &          & \multicolumn{1}{c}{$cis$ }                 & \multicolumn{1}{c}{TS}           & \multicolumn{1}{c}{$trans$}     && \multicolumn{1}{c}{$cis$}          & \multicolumn{1}{c}{TS}           & \multicolumn{1}{c}{$trans$}    \\    \cmidrule{4-6} \cmidrule{8-10}
         &                        &          &                      &              &                \\
B97-1$^a$& $E_\mathrm{eq}$        & / $E_h$  & -499.2369962 & -499.2233235 & -499.2339542 &  & -499.2386084 & -499.2243882 & -499.2398478 \\
         & $E_\mathrm{ZPE}$       & / $E_h$  & -499.080972  & -499.06842   & -499.078206  &  & -499.082942  & -499.069577  & -499.084122  \\
         & $\omega$               & /\si{\wn}& 85.6         & 158.6        & 84.4         &  & 86.3         & 173.9        & 78.8         \\
         & $\Delta E_\mathrm{eq}$ & / \si{\kJ\per\mol} & 0            & 35.9         & 8.0          &  & 3.3          & 40.6         & 0            \\
         &                        & /\si{\wn}& 0            & 3001         & 668          &  & 272          & 3393         & 0            \\
         & $\Delta E_\mathrm{ZPE}$& / \si{\kJ\per\mol} & 0            & 32.0         & 7.3          &  & 3.1          & 37.1         & 0            \\
         &                        & /\si{\wn}& 0            & 2676         & 607          &  & 259          & 3105         & 0            \\ \midrule
MP2$^b$  & $E_\mathrm{eq}$        & / $E_h$  & -498.1872922 & -498.1752513 & -498.1841598 &  & -498.1880392 & -498.1751578 & -498.1893045 \\
         & $E_\mathrm{ZPE}$       & / $E_h$  & -498.030536  & -498.01949   & -498.027667  &  & -498.031632  & -498.019528  & -498.032875  \\
         & $\omega$               & /\si{\wn}& 79.3         & 145.3        & 65.7         &  & 83.7         & 162.9        & 77           \\
         & $\Delta E_\mathrm{eq}$ & / \si{\kJ\per\mol} & 0            & 31.6         & 8.2          &  & 3.3          & 37.1         & 0            \\
         &                        & /\si{\wn}& 0            & 2643         & 687          &  & 278          & 3105         & 0            \\
         & $\Delta E_\mathrm{ZPE}$& / \si{\kJ\per\mol} & 0            & 28.1         & 7.5          &  & 3.3          & 34.1         & 0            \\
         &                        & /\si{\wn}& 0            & 2352         & 630          &  & 273          & 2848         & 0            \\ \midrule
hybrid$^c$& $E_\mathrm{ZPE}$       & / $E_h$ & -498.031268  & -498.0203478 & -498.0284116 &  & -498.0323728 & -498.0203466 & -498.0335787 \\
         & $\Delta E_\mathrm{ZPE}$& / \si{\kJ\per\mol} & 0            & 27.7         & 7.5          &  & 3.2          & 33.7         & 0            \\
         &                        & /\si{\wn}& 0            & 2317         & 627          &  & 265          & 2817         & 0            \\
         & Pop.$^e$             & / \%       & 95           &              & 5            &  & 22           &              & 78           \\  
         & $\Delta E_{12}\,^f$& / \si{\kJ\per\mol}     &    6.1    &     --          &    --       &&         --    &   --     &      0                             \\
         &                        & /\si{\wn}&    507    &        --        &    --        &&      --       &  --       &   0                              \\
         \bottomrule
    \end{tabular}
    
    \smallskip
    \begin{minipage}{0.8\textwidth}
    $^a$ B97-1/cc-pVTZ anharmonic calculation \hspace{5mm} 
    $^b$ MP2/cc-pVTZ anharmonic calculation  \hspace{5mm} 
    $^c$ hybrid B97-1/MP2/cc-pVTZ calculation, see text \hspace{5mm} 
    $^d$ For the transition states, $\Delta E_\mathrm{ZPE}$ is calculated according to the formula $\Delta E_\mathrm{ZPE} = E(\mathrm{TS})- E(\mathrm{GS}) - \omega(\mathrm{TS})/2$ \hspace{5mm} 
    $^e$ Population at 300 K \hspace{5mm} 
    $^f$ Energy difference between the lower state conformers of 1- and 2-naphthaldehyde
    \end{minipage}
\end{sidewaystable}

\begin{table}[htbp]
\caption{Energies in \si{\kJ\per\mol} of the four conformers of naphthaldehyde relative to the most stable isomer \textit{trans} 2-naphthaldehyde computed at various levels of theory. CBS values are extrapolated from aug-cc-pVTZ and aug-cc-pVQZ basis set calculations.}
\label{tab:QM-comparison}
\begin{minipage}{\linewidth}
\begin{tabular}{ll*{2}Sl*{2}S}
\toprule
Method &&\multicolumn{2}{c}{1-naphthaldehyde} && \multicolumn{2}{c}{2-naphthaldehyde}\\
\cmidrule{3-4}\cmidrule{6-7}
&& {\textit{cis}} & {\textit{trans}} && {\textit{cis}} & {\textbf{\textit{trans}}}\\
\cmidrule{1-1}\cmidrule{3-4}\cmidrule{6-7}
MP2/cc-pVTZ\footnote{See values reported in Table~\ref{tab:EcalcAL}.}       && 5.3 & 13.5 && 3.3 & 0.0\\
MP2/CBS          && 5.8 & 13.9 && 3.6 & 0.0\\
CCSD(T)/CBS      && 6.6 & 14.0 && 3.6 & 0.0\\
PNO-LCCSD(T)-F12/aug-cc-pVTZ && 6.7 & 13.8 && 3.5 & 0.0\\
\bottomrule
\end{tabular}
\end{minipage}
\end{table}

\begin{table*}[htbp]
  \centering \scriptsize
  \caption{Calculated IR fundamental vibrational modes of $cis$-1-naphthaldehyde and $trans$-2-naphthaldehyde and comparison with proposed experimental assignments. Energies are given in \si{\wn} intensities in \si{\km\per\mol}.}
    \begin{tabular}{lc rrcrd{2.1}d{3.0}r c rrcrd{2.1}d{3.0}r}
    \toprule
        &  & \multicolumn{7}{c}{$cis$-1-naphthaldehyde} && \multicolumn{7}{c}{$trans$-2-naphthaldehyde}\\ \cmidrule{3-9} \cmidrule{11-17}
        &  & \multicolumn{2}{c}{B97-1}  && \multicolumn{2}{c}{MP2} & \multicolumn{1}{c}{Hybrid}  &  \multicolumn{1}{c}{Exp.} && \multicolumn{2}{c}{B97-1}  && \multicolumn{2}{c}{MP2} & \multicolumn{1}{c}{Hybrid}  &  \multicolumn{1}{c}{Exp.} \\ \cmidrule{3-4}\cmidrule{6-7} \cmidrule{11-12} \cmidrule{14-15}
    $\nu$ & sym. & \multicolumn{1}{c}{$E_\mathrm{h}$} & \multicolumn{1}{c}{$E_\mathrm{a}$} && \multicolumn{1}{c}{$E_\mathrm{h}$} & \multicolumn{1}{c}{$I_\mathrm{h}$} & \multicolumn{1}{c}{$E$} & \multicolumn{1}{c}{$E$}& & \multicolumn{1}{c}{$E_\mathrm{h}$} & \multicolumn{1}{c}{$E_\mathrm{a}$} && \multicolumn{1}{c}{$E_\mathrm{h}$} & \multicolumn{1}{c}{$I_\mathrm{h}$} & \multicolumn{1}{c}{$E$} & \multicolumn{1}{c}{$E$}     \\
    \midrule
1  & A'  & 3224 & 3070 && 3268 & 7.5   & 3114 & 3064  &  & 3189 & 3062 && 3234 & 3.5   & 3107 &      \\
2  & A'  & 3184 & 3062 && 3232 & 8.5   & 3109 & 3064  &  & 3185 & 3061 && 3233 & 11.6  & 3109 & 3070 \\
3  & A'  & 3182 & 3077 && 3230 & 14.0  & 3125 & 3064  &  & 3173 & 3052 && 3220 & 13.4  & 3099 & 2980 \\
4  & A'  & 3167 & 3032 && 3215 & 10.5  & 3080 & 3064  &  & 3161 & 3045 && 3205 & 0.7   & 3089 &      \\
5  & A'  & 3158 & 3055 && 3202 & 6.4   & 3099 & 3064  &  & 3159 & 3009 && 3203 & 5.3   & 3053 &      \\
6  & A'  & 3154 & 3009 && 3200 & 0.4   & 3055 & 3064  &  & 3155 & 3013 && 3200 & 2.3   & 3058 &      \\
7  & A'  & 3150 & 3018 && 3195 & 2.7   & 3064 & 3064  &  & 3143 & 2998 && 3184 & 6.1   & 3039 &      \\
8  & A'  & 2861 & 2684 && 2936 & 118.0 & 2759 & 2721  &  & 2870 & 2680 && 2949 & 108.4 & 2760 & 2717 \\
9  & A'  & 1772 & 1743 && 1740 & 161.0 & 1712 & 1715  &  & 1781 & 1752 && 1745 & 231.6 & 1717 & 1720 \\
10 & A'  & 1653 & 1611 && 1669 & 3.7   & 1628 &       &  & 1660 & 1620 && 1678 & 6.0   & 1638 & 1637 \\
11 & A'  & 1626 & 1592 && 1625 & 7.0   & 1591 & 1601  &  & 1635 & 1599 && 1635 & 4.5   & 1600 & 1603 \\
12 & A'  & 1605 & 1566 && 1619 & 11.1  & 1580 & 1583  &  & 1606 & 1572 && 1617 & 5.2   & 1582 &      \\
13 & A'  & 1542 & 1504 && 1561 & 29.8  & 1523 & 1516  &  & 1539 & 1504 && 1556 & 2.2   & 1521 &      \\
14 & A'  & 1488 & 1456 && 1495 & 0.9   & 1462 &       &  & 1495 & 1467 && 1506 & 10.7  & 1478 & 1468 \\
15 & A'  & 1474 & 1437 && 1491 & 7.3   & 1454 & 1457  &  & 1471 & 1439 && 1491 & 12.7  & 1459 & 1447 \\
16 & A'  & 1438 & 1405 && 1481 & 5.2   & 1448 & 1457  &  & 1424 & 1408 && 1472 & 2.5   & 1456 &      \\
17 & A'  & 1418 & 1378 && 1443 & 2.9   & 1404 &       &  & 1401 & 1363 && 1456 & 0.1   & 1418 &      \\
18 & A'  & 1388 & 1351 && 1435 & 2.8   & 1399 &       &  & 1388 & 1357 && 1416 & 1.6   & 1385 &      \\
19 & A'  & 1368 & 1335 && 1414 & 2.6   & 1381 &       &  & 1369 & 1338 && 1374 & 25.2  & 1343 & 1347 \\
20 & A'  & 1292 & 1249 && 1290 & 2.1   & 1248 &       &  & 1285 & 1271 && 1294 & 37.5  & 1279 & 1262 \\
21 & A'  & 1237 & 1211 && 1266 & 18.6  & 1240 & 1220? &  & 1278 & 1255 && 1284 & 4.5   & 1261 &      \\
22 & A'  & 1232 & 1204 && 1240 & 16.5  & 1212 & 1214? &  & 1236 & 1212 && 1259 & 0.0   & 1235 &      \\
23 & A'  & 1189 & 1165 && 1192 & 12.7  & 1168 & 1171  &  & 1184 & 1163 && 1195 & 41.5  & 1174 & 1154 \\
24 & A'  & 1184 & 1155 && 1181 & 2.1   & 1152 &       &  & 1172 & 1154 && 1178 & 0.1   & 1160 &      \\
25 & A'  & 1165 & 1142 && 1166 & 3.4   & 1143 &       &  & 1165 & 1166 && 1161 & 0.1   & 1163 &      \\
26 & A'  & 1097 & 1077 && 1103 & 2.3   & 1084 &       &  & 1137 & 1124 && 1136 & 21.6  & 1123 & 1121 \\
27 & A'  & 1069 & 1041 && 1074 & 35.4  & 1046 & 1056  &  & 1039 & 1022 && 1047 & 2.3   & 1030 &      \\
28 & A'  & 1042 & 1012 && 1050 & 3.9   & 1020 &       &  & 969  & 956  && 970  & 1.0   & 956  &      \\
29 & A'  & 895  & 881  && 893  & 15.3  & 878  & 887   &  & 891  & 879  && 893  & 6.2   & 881  & 882  \\
30 & A'  & 805  & 795  && 807  & 0.7   & 797  &       &  & 785  & 774  && 789  & 21.2  & 778  & 785  \\
31 & A'  & 718  & 704  && 718  & 15.1  & 704  & 712   &  & 766  & 755  && 767  & 23.7  & 757  & 763  \\
32 & A'  & 659  & 649  && 656  & 18.3  & 646  & 647   &  & 640  & 633  && 632  & 1.9   & 626  &      \\
33 & A'  & 553  & 547  && 551  & 2.5   & 544  & 548   &  & 612  & 606  && 609  & 3.6   & 603  & 603  \\
34 & A'  & 502  & 495  && 498  & 1.6   & 492  & 497   &  & 520  & 514  && 516  & 0.1   & 511  &      \\
35 & A'  & 437  & 433  && 434  & 1.5   & 431  & 430   &  & 390  & 389  && 386  & 5.5   & 384  & 388  \\
36 & A'  & 366  & 362  && 360  & 0.1   & 356  &       &  & 349  & 350  && 348  & 2.8   & 349  &      \\
37 & A'  & 218  & 166  && 219  & 6.5   & 166  & 225 ? &  & 177  & 184  && 175  & 6.0   & 183  & 177  \\
\\
38 & A'' & 1022 & 1003 && 1015 & 0.4   & 995  &       &  & 1028 & 1011 && 1018 & 1.1   & 1001 &      \\
39 & A'' & 1016 & 1011 && 977  & 0.7   & 973  &       &  & 1002 & 994  && 968  & 0.1   & 961  &      \\
40 & A'' & 995  & 994  && 962  & 0.8   & 962  &       &  & 998  & 995  && 960  & 0.1   & 957  &      \\
41 & A'' & 979  & 975  && 951  & 0.1   & 947  &       &  & 972  & 968  && 948  & 1.5   & 943  &      \\
42 & A'' & 937  & 930  && 909  & 0.0   & 902  &       &  & 918  & 909  && 878  & 9.7   & 869  & ?    \\
43 & A'' & 893  & 880  && 870  & 0.0   & 856  &       &  & 880  & 864  && 859  & 20.1  & 843  & 857  \\
44 & A'' & 823  & 818  && 806  & 91.7  & 801  & 802   &  & 843  & 831  && 835  & 35.4  & 822  & 820  \\
45 & A'' & 792  & 777  && 755  & 8.0   & 740  &       &  & 788  & 778  && 754  & 27.7  & 745  & 742  \\
46 & A'' & 753  & 740  && 725  & 0.3   & 711  &       &  & 760  & 751  && 726  & 0.2   & 717  &      \\
47 & A'' & 639  & 637  && 614  & 0.0   & 612  &       &  & 645  & 640  && 620  & 0.9   & 614  &      \\
48 & A'' & 539  & 529  && 529  & 6.0   & 520  & 525   &  & 512  & 508  && 502  & 0.3   & 497  &      \\
49 & A'' & 479  & 472  && 469  & 0.4   & 462  &       &  & 485  & 479  && 477  & 15.9  & 471  & 472  \\
50 & A'' & 414  & 408  && 407  & 3.6   & 402  & 400   &  & 401  & 395  && 393  & 0.0   & 387  &      \\
51 & A'' & 273  & 266  && 265  & 5.7   & 258  & 255   &  & 297  & 289  && 293  & 3.3   & 285  &      \\
52 & A'' & 181  & 178  && 178  & 0.9   & 175  &       &  & 192  & 188  && 187  & 4.7   & 184  & 181  \\
53 & A'' & 152  & 146  && 147  & 3.7   & 141  & 143   &  & 160  & 156  && 157  & 4.8   & 152  & 152  \\
54 & A'' & 86   & 76   && 79   & 3.5   & 70   & 76    &  & 79   & 78   && 77   & 1.6   & 76   &      \\
    \bottomrule
    \end{tabular}%
    
    \medskip
    \begin{minipage}{0.9 \textwidth}
    \textbf{Note.} $h$ in $E_h$ stands for harmonic, and $a$ in $E_a$ for anharmonic
    \end{minipage}
      \label{tab:NaphthalVibFreqTot}%
\end{table*}%

\begin{table}[ht!]
    \centering 
    \begin{tabular}{ll rrrrr}
    \toprule
           &      & MP2   & & hybrid$^b$    & \multicolumn{1}{c}{exp.}         & \multicolumn{1}{c}{$\delta_\mathrm{hybrid}\ ^c$} \\
           &      & \multicolumn{1}{r}{$B_e$ } &   \multicolumn{1}{c}{$\Delta^a$}           &  \multicolumn{1}{c}{$B_0$}   &      \multicolumn{1}{c}{$B_0$}          & / \%                                              \\ \midrule
\multicolumn{3}{l}{\textbf{$trans$-1-hydroxynaphthalene}} &       &                             \\
$A$      & / \si{MHz}&  1947  & -14 & 1934 & 1942.100623\,(53)  & 0.42  \\
$B$      & / \si{MHz}&  1140  & -8  & 1132 & 1133.623211\,(21)  & 0.14  \\
$C$      & / \si{MHz}&  719    & -5  & 714   &  716.017762\,(14)  & 0.28  \\
           &      &       &     &       &         &                     \\
\multicolumn{3}{l}{\textbf{$cis$-2-hydroxynaphthalene}}      &       &                            \\
$A$      & / \si{MHz}&  2862 & -25 & 2837 & 2849.155543\,(64)  & 0.43  \\
$B$      & / \si{MHz}&  828   & -4  & 823   &  824.632161\,(25)  & 0.20  \\
$C$      & / \si{MHz}&  642   & -4  & 638   &  639.723697\,(15)  & 0.27  \\
\\
\multicolumn{3}{l}{\textbf{$cis$-1-naphthaldehyde}}      &       &                                \\
$A$      & / \si{MHz}& 1389 & -8 &  1380 &   1384.40872\,(21) &  0.32                   \\
$B$      & / \si{MHz}& 1004 & -8 &  995   &   999.015243\,(71) &  0.40                   \\
$C$      & / \si{MHz}&  583  & -4 &  579   &   580.551953\,(19) &  0.27                   \\
           &      &       &     &       &         &                    \\
\multicolumn{3}{l}{\textbf{$trans$-2-naphthaldehyde}}      &       &                                \\
$A$      & / \si{MHz} & 2824 & -22  &  2802  & 2810.43194\,(15)  & 0.30        \\
$B$      & / \si{MHz} & 583   & -3   &  580    & 581.095981\,(24)  & 0.19        \\
$C$      & / \si{MHz} & 483   & -3   &  480    & 481.700564\,(18)  & 0.35        \\
\bottomrule
\end{tabular}

    \medskip
    \begin{minipage}{0.7 \textwidth}
    $^a \Delta= B_0 - B_e$ \hspace{5mm}
    $^b B_{0}^{hybrid} = B_{e}^{MP2} - (B_{e}^{B97-1} - B_{0}^{B97-1})$ \\
    $^c \delta_\mathrm{hybrid} = (B_{\mathrm{exp.}}-B_{\mathrm{hybrid}}) / B_{\mathrm{hybrid}} \times 100$ 
    \end{minipage}
    \caption{Rotational constants of the four studied species at various stages of the calculations and comparison with the experimental values. }
    \label{tab:RotCtsGS}
\end{table}

\begin{table}[ht!]
    \centering 
    \begin{tabular}{ll rrrrrrr}
    \toprule
           &      & MP2   & $\Delta^a$ & hybrid$^b$    & scaled$^c$ & exp.         & $\delta_\mathrm{hybrid}\ ^d$& $\delta_\mathrm{scaled}\ ^e$ \\
           &      &   &              &     &          &        & / \%                    & / \%                           \\ \midrule
\multicolumn{2}{l}{\textbf{Equilibrium}}      &       &     &       &         &             &      &        \\
$A_e$      & / \si{MHz}& 1947 &     &       &         &             &      &        \\
$B_e$      & / \si{MHz}& 1140 &     &       &         &             &      &        \\
$C_e$      & / \si{MHz}& 719   &     &       &         &             &      &        \\
           &      &       &     &       &         &             &      &        \\
\multicolumn{2}{l}{\textbf{Ground state}}      &       &     &       &         &             &      &        \\
$A_0$      & / \si{MHz}&       &     & 1934  &         & 1942.100623\,(53)   & 0.42 &        \\
$B_0$      & / \si{MHz}&       &     & 1132  &         & 1133.623211\,(21)   & 0.16 &        \\
$C_0$      & / \si{MHz}&       &     & 714   &         &  716.017762\,(14)   & 0.26 &        \\
           &      &       &     &       &         &             &      &        \\
\multicolumn{2}{l}{$\bf v_{51}=1$} &       &     &       &         &             &      &        \\
$E$        & /\si{\wn}&       &     & 136   &         &             &      &        \\
$A_v$      & / \si{MHz}&       & -14 & 1933 & 1941    &  1941.73727\,(53) & 0.44 & 0.015  \\
$B_v$      & / \si{MHz}&       & -8  & 1132 & 1134    & 1132.951050\,(68) & 0.11 & -0.051 \\
$C_v$      & / \si{MHz}&       & -4  & 715   & 717     &  716.402565\,(43) & 0.23 & -0.030 \\
           &      &       &     &       &         &             &      &        \\
\multicolumn{2}{l}{$\bf v_{50}=1$}      &       &     &       &         &             &      &        \\
$E$        & /\si{\wn}&       &     & 166.4 &         &             &      &        \\
$A_v$      & / \si{MHz}&       & -16 & 1931 & 1939    &  1939.60653\,(22)  & 0.44 & 0.016  \\
$B_v$      & / \si{MHz}&       & -6  & 1133 & 1135    & 1134.394385\,(36)  & 0.11 & -0.053 \\
$C_v$      & / \si{MHz}&       & -4  & 715   & 717     &  716.571336\,(25)  & 0.23 & -0.032 \\
           &      &       &     &       &         &             &      &        \\
\multicolumn{2}{l}{$\bf v_{51}=2$}     &       &     &       &         &             &      &        \\
$E$        & /\si{\wn}&       &     & 272.2 &         &             &      &        \\
$A_v$      & / \si{MHz}&       & -15 & 1933 & 1941    &  1941.1817\,(43) & 0.43 & 0.003  \\
$B_v$      & / \si{MHz}&       & -9  & 1131 & 1133    &  1132.3400\,(16) & 0.12 & -0.042 \\
$C_v$      & / \si{MHz}&       & -4  & 715   & 717     & 716.785770\,(55) & 0.23 & -0.027 \\
           &      &       &     &       &         &             &      &        \\
\multicolumn{2}{l}{$\bf v_{50}=2$}      &       &     &       &         &             &      &        \\
$E$        & /\si{\wn}&       &     & 332.8 &         &             &      &        \\
$A_v$      & / \si{MHz}&       & -19 & 1929 & 1937    &  1937.1523\,(11) & 0.44 & 0.019  \\
$B_v$      & / \si{MHz}&       & -6  & 1134 & 1136    & 1135.16069\,(49) & 0.11 & -0.052 \\
$C_v$      & / \si{MHz}&       & -3  & 715   & 717     & 717.121942\,(55) & 0.23 & -0.030 \\
\bottomrule
\end{tabular}

    \medskip
    \begin{minipage}{0.7 \textwidth}
    $^a \Delta= B_0 - B_e$ \hspace{5mm}
    $^b B_{0}^{hybrid} = B_{e}^{MP2} - (B_{e}^{B97-1} - B_{0}^{B97-1})$ \\
    $^c B_{\mathrm{scaled}}= B_{v\ \mathrm{hybrid}} \times B_{0\ \mathrm{exp}}/ B_{0\ \mathrm{hybrid}}$ \\
    $^d \delta_\mathrm{hybrid} = (B_{\mathrm{exp.}}-B_{\mathrm{hybrid}}) / B_{\mathrm{hybrid}} \times 100$ \\
    $^e \delta_\mathrm{scaled} = (B_{\mathrm{exp.}}-B_{\mathrm{scaled}}) / B_{\mathrm{scaled}} \times 100$ 
    \end{minipage}
    \caption{Rotational constants of $trans$-1-hydroxynaphthalene in the ground and excited vibrational states. Error ($1\sigma$) on experimental parameters are given between parenthesis in units of the last digit. }
    \label{tab:RotCtsES}
\end{table}

\begin{table*}[htbp]
  \centering \small
  \caption[Full set of rotational parameters (in \si{MHz}) for $trans$-1-hydroxynaphthalene in excited vibrational states]{Full set of rotational parameters (in \si{MHz}) for $trans$-1-hydroxynaphthalene in excited vibrational states. Error ($1\sigma$) on parameters are given between parenthesis in units of the last digit. Parameters in brackets are fixed to the ground state value.}
    \begin{tabular}{l d{4.10}d{4.10}d{4.10}d{4.10}}
    \toprule
          & \multicolumn{1}{c}{$v_{51}=1$} &    \multicolumn{1}{c}{$v_{50}=1$}&    \multicolumn{1}{c}{$v_{51}=2$} & \multicolumn{1}{c}{$v_{50}=2$}\\ \midrule
    $A_v$                       &   1941.73727\,(53)  &  1939.60653\,(22) &  1941.1817\,(43) &     1937.1523\,(11) \\
    $B_v$                       &  1132.951050\,(68)  & 1134.394385\,(36) &  1132.3400\,(16) &    1135.16069\,(49) \\
    $C_v$                       &   716.402565\,(43)  &  716.571336\,(25) & 716.785770\,(55) &    717.121942\,(55) \\
    $\Delta_J    \times 10^{6}$ &      17.9061\,(20)  &     18.0313\,(10) &    17.9912\,(16) &       18.0842\,(16) \\
    $\Delta_{JK} \times 10^{3}$ &    0.0185871\,(92)  &   0.0134017\,(43) &   0.016961\,(48) &      0.014049\,(34) \\
    $\Delta_K    \times 10^{3}$ &      0.04257\,(14)  &    0.045458\,(62) &    0.04502\,(35) &       0.03692\,(23) \\
    $\delta_J    \times 10^{6}$ &      6.41897\,(11)  &     6.47433\,(58) &    [6.43578] &     [6.43578] \\
    $\delta_K    \times 10^{3}$ &    0.0360674\,(88)  &   0.0338925\,(51) &    [0.0351740] &     [0.0351740] \\
 \\
    \# lines  & \multicolumn{1}{c}{2 567} &\multicolumn{1}{c}{2 547}  &  \multicolumn{1}{c}{1 093}  &    \multicolumn{1}{c}{1 035}    \\
    RMS / kHz & \multicolumn{1}{c}{34}& \multicolumn{1}{c}{34}    &  \multicolumn{1}{c}{72}           & \multicolumn{1}{c}{44}      \\
    \bottomrule
    \end{tabular}%
    
    \medskip
    \begin{minipage}{0.8 \textwidth}
    $^a$ Hybrid rotational constants and MP2 centrifugal distorsion terms, see text\\
    $^b \delta = (B_{\mathrm{Exp.}}-B_{\mathrm{Calc}}) / B_{\mathrm{Calc.}} \times 100$
    \end{minipage}
  \label{tab:RotCtsTot_1oh}%
\end{table*}%

\begin{table}[ht!]
    \centering 
    \caption{Calculated charges on atoms in naphthalene, benzaldehyde and the two isomers of naphthaldehyde at the MP2/aug-cc-pVQZ level using NBO6 software. Values are given in atomic units. See Figure~5 for atomic numbering.}
  \begin{tabular}{lr rr rr r rrr}
    \toprule
&  & \multicolumn{2}{c}{1-naphthaldehyde} & \multicolumn{2}{c}{2-naphthaldehyde} & \multicolumn{1}{c}{naphthalene} & \multicolumn{3}{c}{benzaldehyde}\\
atom    &    \#    &    \multicolumn{1}{c}{\textit{cis}}    &    \multicolumn{1}{c}{\textit{trans}}    &    \multicolumn{1}{c}{\textit{cis}}    &    \multicolumn{1}{c}{\textit{trans}}    &       &  atom    &    \# \\ \midrule
C	&	1	&	-0.0738	&	-0.0937	&	-0.1125	&	-0.0997	&	-0.1726	&	C	&	1	&	-0.0954	\\
C	&	2	&	-0.2295	&	-0.2246	&	-0.1247	&	-0.1105	&	-0.2177	&	C	&	2	&	-0.2109	\\
C	&	3	&	-0.1628	&	-0.1388	&	-0.2209	&	-0.2056	&	-0.2177	&	C	&	3	&	-0.2029	\\
C	&	4	&	-0.1641	&	-0.2116	&	-0.1730	&	-0.1916	&	-0.1726	&	C	&	4	&	-0.1497	\\
C	&	5	&	-0.1352	&	-0.1279	&	-0.2285	&	-0.1341	&	-0.1726	&	C	&	5	&	-0.2112	\\
C	&	6	&	-0.2326	&	-0.2358	&	-0.1421	&	-0.2395	&	-0.2177	&	C	&	6	&	-0.2130	\\
C	&	7	&	-0.1812	&	-0.2067	&	-0.2324	&	-0.2096	&	-0.2177	&	H	&	7	&	0.2354	\\
C	&	8	&	-0.2247	&	-0.2101	&	-0.1789	&	-0.1277	&	-0.1726	&	H	&	8	&	0.2142	\\
C	&	9	&	0.0053	&	0.0015	&	-0.0601	&	-0.1527	&	-0.0192	&	H	&	9	&	0.2068	\\
C	&	10	&	-0.0863	&	-0.0360	&	-0.0154	&	0.0086	&	-0.0192	&	H	&	10	&	0.2141	\\
H	&	11	&	0.2483	&	0.2108	&	0.2132	&	0.2020	&	0.1928	&	H	&	11	&	0.2134	\\
H	&	12	&	0.2132	&	0.2165	&	0.2168	&	0.2167	&	0.2072	&	H	&	12	&	0.1093	\\
H	&	13	&	0.2158	&	0.2151	&	0.2031	&	0.2152	&	0.2072	&	O	&	13	&	-0.4800	\\
H	&	14	&	0.2002	&	0.1969	&	0.2132	&	0.1996	&	0.1928	&	C	&	14	&	0.3698	\\
H	&	15	&	0.2087	&	0.2088	&	0.2128	&	0.2105	&	0.1928	&		&		&		\\
H	&	16	&	0.2074	&	0.2050	&	0.2160	&	0.2335	&	0.2072	&		&		&		\\
H	&	17	&	0.2145	&	0.2366	&	0.2166	&	0.1975	&	0.2072	&		&		&		\\
H	&	18	&	0.1087	&	0.1136	&	0.1095	&	0.1082	&	0.1928	&		&		&		\\
O	&	19	&	-0.4903	&	-0.4805	&	-0.4806	&	-0.4817	&		&		&		&		\\
C	&	20	&	0.3583	&	0.3607	&	0.3677	&	0.3607	&		&		&		&		\\
  \bottomrule
    \end{tabular}                                                       
    \label{tab:Qal}
\end{table}

\begin{table}[ht!]
    \centering 
    \caption{Calculated charges on atoms in naphthalene, hydroxybenzene, and the two isomers of hydroxynaphthalene at the MP2/aug-cc-pVQZ level using NBO6 software. Values are given in atomic units. See Figure~5 for atomic numbering.}
  \begin{tabular}{lr rr rr r  rrr}
    \toprule
&  & \multicolumn{2}{c}{1-hydroxynaphthalene} & \multicolumn{2}{c}{2-hydroxynaphthalene} &  \multicolumn{1}{c}{naphthalene} & \multicolumn{3}{c}{hydroxybenzene}\\
  atom    &    \#    &  \multicolumn{1}{c}{\textit{cis}}    &    \multicolumn{1}{c}{\textit{trans}}    &    \multicolumn{1}{c}{\textit{cis}}    &    \multicolumn{1}{c}{\textit{trans}}    &        & atom    &    \# \\ \midrule
&	1	&	0.2765	&	0.2716	&	-0.3345	&	-0.2098	&	-0.1726	&	C	&	1	&	0.3414	\\
&	2	&	-0.2756	&	-0.2554	&	0.3487	&	0.2520	&	-0.2177	&	C	&	2	&	-0.2838	\\
&	3	&	-0.2136	&	-0.2243	&	-0.2971	&	-0.2815	&	-0.2177	&	C	&	3	&	-0.1891	\\
&	4	&	-0.1675	&	-0.1637	&	-0.1172	&	-0.1845	&	-0.1726	&	C	&	4	&	-0.2452	\\
&	5	&	-0.1778	&	-0.1583	&	-0.2127	&	-0.1457	&	-0.1726	&	C	&	5	&	-0.1900	\\
&	6	&	-0.2466	&	-0.2540	&	-0.2294	&	-0.2481	&	-0.2177	&	C	&	6	&	-0.3103	\\
&	7	&	-0.1938	&	-0.1356	&	-0.2295	&	-0.2186	&	-0.2177	&	H	&	7	&	0.2241	\\
&	8	&	-0.1811	&	-0.2261	&	-0.1486	&	-0.1936	&	-0.1726	&	H	&	8	&	0.2142	\\
&	9	&	-0.0945	&	-0.0539	&	0.0058	&	-0.0029	&	-0.0192	&	H	&	9	&	0.2142	\\
&	10	&	-0.0007	&	-0.0876	&	-0.0411	&	-0.0441	&	-0.0192	&	H	&	10	&	0.2140	\\
&	11	&	0.1795	&	0.2282	&	0.1936	&	0.2048	&	0.1928	&	H	&	11	&	0.2077	\\
&	12	&	0.2105	&	0.2033	&	0.2099	&	0.2134	&	0.2072	&	H	&	12	&	0.4729	\\
&	13	&	0.2161	&	0.2118	&	0.2144	&	0.2119	&	0.2072	&	O	&	13	&	-0.6700	\\
&	14	&	0.2070	&	0.2075	&	0.2065	&	0.1990	&	0.1928	&		&		&		\\
&	15	&	0.1963	&	0.2026	&	0.1960	&	0.2069	&	0.1928	&		&		&		\\
&	16	&	0.2254	&	0.2026	&	0.2267	&	0.2104	&	0.2072	&		&		&		\\
&	17	&	0.2146	&	0.2101	&	0.2046	&	0.2099	&	0.2072	&		&		&		\\
&	18	&	0.4751	&	0.4785	&	0.4742	&	0.4749	&	0.1928	&		&		&		\\
&	19	&	-0.6498	&	-0.6573	&	-0.6702	&	-0.6542	&		&		&		&		\\
  \bottomrule
    \end{tabular}                                                        
    \label{tab:Qol}
\end{table}